\documentclass{ptephy} 
\usepackage{amsmath}
\usepackage{amsfonts}
\usepackage{graphicx}
\def\slash#1{\not\!#1}
\newcommand{\cal}{\mathcal}

\begin{document}
\title{Pion properties at finite nuclear density based on in-medium chiral perturbation theory}
\author{
  \name{\fname{Soichiro} \surname{Goda}}{1,\ast},
  \name{\fname{Daisuke} \surname{Jido}}{2}}
\address{
  \affil{1}{Department of Physics, Graduate School of Science, Kyoto University Kyoto 606-8502, Japan}
  \affil{2}{Department of Physics, Tokyo Metropolitan University, Hachioji, Tokyo 192-0397, Japan}
\email{gouda@ruby.scphys.kyoto-u.ac.jp}}

\begin{abstract}
The in-medium pion properties, {\it i.e.} the temporal pion decay constant $f_t$, the pion mass $m_\pi^*$ and the wave function renormalization, in symmetric nuclear matter are calculated in an in-medium chiral perturbation theory up to the next-to-leading order of the density expansion $O(k_F^4)$.
The chiral Lagrangian for the pion-nucleon interaction is determined in vacuum, and the low energy constants are fixed by the experimental observables. 
We carefully define the in-medium state of pion and find that the pion wave function renormalization plays an essential role for the in-medium pion properties. 
We show that the linear density correction is dominant and the next-to-leading corrections are not so large at the saturation density, while their contributions can be significant in higher densities.
The main contribution of the next-to-leading order comes from the double scattering term.
We also discuss whether the low energy theorems, the Gell-Mann--Oakes--Renner relation and the Glashow--Weinberg relation, are satisfied in nuclear medium beyond the linear density approximation.
We also find that the wave function renormalization is enhanced as largely as 50\% at the saturation density including the next-to-leading contribution and the wave function renormalization could be measured in the in-medium $\pi^0\to \gamma\gamma$ decay.
\end{abstract}
\subjectindex{D33}
\maketitle

\section{Introduction}
Spontaneous breakdown of chiral symmetry ($ \chi $SSB) $ SU(N_f)_L \times SU(N_f)_R \to SU(N_f)_V $ characterizes the vacuum and low-energy dynamics of Quantum ChromoDynamics (QCD)~\cite{Nam61}. 
The non-vanishing chiral condensate $ \langle \bar{q} q \rangle $ is considered as one of the order parameters of $\chi$SSB and gives a characteristic scale for hadron physics.
$\chi$SSB is considered to be responsible for the origin of constituent quark mass after the current quark mass is given by the Higgs condensate slightly.
According to the spontaneous breakdown, the pseudoscalar mesons such as $\pi , K , \eta$ appear as the Nambu-Goldstone (NG) bosons.

Recently, in order to investigate the mechanism of the dynamical mass generation, partial restoration of the chiral symmetry in the nuclear medium has gained considerable attention.
This phenomenon is incomplete restoration of chiral symmetry with sufficient reduction of the absolute value of the chiral condensate in the medium and will lead to various changes of hadron properties.
Once we understand the partial restoration of chiral symmetry, we can predict other in-medium hadronic quantities  through low energy theorems and vice versa.

From this point of view, vast theoretical and experimental efforts are devoted for this topic.
The density dependence of the chiral condensate is evaluated in various approaches, for example, the well-known linear density approximation providing the model-independent low density theorem \cite{Coh91,Dru91}, the relativistic Bruckner-Hartree-Fock theory approach\cite{Bro96} and systematic calculations by in-medium chiral perturbation theory beyond the linear density calculation \cite{Mei02,Kai08}.
According to the model-independent linear density approximation, the leading density correction to the chiral condensate is determined by the $ \pi $N sigma term $ \sigma_{\pi N}$ and with the empirical value of the sigma term it has been found that the leading correction gives enough large contribution at the normal nuclear density:
\begin{equation}
\frac{\langle \bar{q} q \rangle^*}{ \langle \bar{q} q \rangle_0 } \approx 1- 0.35 \rho / \rho_0 ,
\end{equation}
where $\langle \bar{q} q \rangle^*$, $\langle \bar{q} q \rangle_0$ is the in-medium and in-vacuum condensate and $\rho_0$ is the normal nuclear density.
Recently, an in-medium sum rule satisfied in any density region has been derived model-independently by current algebra method and the low energy theorems, such as the Gell--Mann--Oakes--Renner relation, the Glashow--Weinberg relation and the Weinberg--Tomozawa relation, are discussed within the linear density order \cite{Jid08}.

To investigate the pion properties in nuclei, deeply bound pionic atoms have received much attention~\cite{Ita00, Gei02, Suz04}.
Theoretically, the binding energies and decay widths of the 1s and 2p 
deeply-bound pionic atom states are estimated \cite{Nat11,Nat11-2,Kol03} and in Refs.~\cite{Gir04,Dor08} hadronic quantities, such as pion optical potential, have been calculated beyond the linear density.
Experimentally the reduction of the chiral condensate is estimated quantitatively through reduction of the $s$-wave isovector parameter $ b_1 $ in the $\pi$-nucleus optical potential~\cite{Suz04}. The $b_{1}$ parameter is regarded as an in-medium isovector $\pi$N scattering length.
These results show that the reduction of $ b_1 $ means the repulsive enhancement of the $s$-wave $\pi$-nucleon interaction in nucleus.
Another examples are low energy $\pi$-nucleus scattering and $\pi \pi $ interaction in nuclei in the scalar-isoscalar channel.
The low energy $\pi$-nucleus scattering also show that s-wave $\pi$-nucleus interaction are enhanced repulsively~\cite{Fri04,Fri05}. 
According to the theoretical discussion given in Refs.~\cite{Hat99,Jid00}, the in-medium $\pi \pi $ interaction in the scalar-isoscalar part will also has attractive enhancement thanks to the partial restoration of chiral symmetry
in nuclear medium and the experimental observation of the invariant mass spectrum of the $\pi\pi$ production off nuclear targets performed in Refs.~\cite{Bon96,Bon00,Cam04} could have a hint of such a enhancement.

In particular the pion decay constant is a fundamental quantity of chiral symmetry breaking.
The in-medium decay constant also has been investigated in the linear density approximation \cite{Tho95}
and recently the chiral condensate and the decay constant have been evaluated in the next-to-leading order based on chiral order counting \cite{Lac10}.
In this paper, we discuss a general in-medium pion state and evaluate in-medium pionic quantities such as the decay constant, mass and the pseudo scalar coupling beyond linear density approximation.
This paper is organized as follows.
In Sec.\ref{In-medium CHPT}, we explain the general formulation of 
the in-medium chiral perturbation theory and discuss an expansion by Fermi momentum counting.
In Sec.\ref{decayconst}, we discuss in-medium pion state and define the in-medium pionic quantities. Here we will find that the pion wave function renormalization plays an important role for the in-medium pionic quantities.
In Sec.\ref{results}, we evaluate the in-medium pion self energy, wave function renormalization, pion decay constant and pseudo-scalar coupling and show the numerical results of the density dependence
of them up to $ O(k_F^4) $ in Fermi momentum expansion in symmetric nuclear matter and in isospin limit.
We also discuss whether the in-medium low energy theorems, the Gell-Mann--Oakes--Renner relation and the Glashow--Weinberg relation, are satisfied or not.
Finally we discuss the in-medium $\pi_0 \to 2 \gamma$ process caused by chiral anomaly.
In Sec.\ref{summary}, we summarize our paper.

\section{In-medium chiral perturbation theory}
\label{In-medium CHPT}
Chiral perturbation theory (CHPT) is a powerful tool describing low energy dynamics of the Nambu-Goldstone (NG) bosons and nucleons as an effective field theory of QCD~\cite{Wei79,Gas84,Gas85,Gas88}.
CHPT is constructed based on chiral symmetry and its spontaneous breaking and consists in systematic expansion of NG boson momentum and the quark mass.
The chiral order counting scheme makes it possible to categorize Lagrangian and Feynman diagrams in terms of powers of momentum and the quark mass and to estimate magnitude of possible corrections for the amplitude, such as the current-current correlation functions.
Thus, CHPT describes quantitatively the S-matrix elements of the QCD currents.

In this decade, chiral effective theory for nuclear matter has been developed~\cite{Oll02,Gir04} and is applied to study nuclear matter properties, such as the nuclear matter energy density\cite{Kai02} and also used to study partial restoration of chiral symmetry and the in-medium changes of the pion properties, such as in-medium pion mass , decay constant \cite{Mei02,Lac10} and 1s and 2p energy levels of deeply-bound pionic atoms \cite{Kol03}.

\subsection{Basics of the formulation}
For the calculations of the in-medium pion quantities, we evaluate the Green's functions in the ground state of nuclear matter, and, in particular, our interests are hadron properties in nuclear medium as a bound state of the nucleon many body system. 
In quantum field theory, the transition amplitude between the ground state in the presence of the external fields is the fundamental quantity, and the generating functional is defined by the transition amplitude, which can be calculated by the path integral formalism of the quantum field theory. In the in-medium chiral perturbation theory, one prepares the ground state of nucleon Fermi gas at asymptotic time as a reference state~\cite{Oll02}.
The generating functional $W$ for the connected Green functions in nuclear matter is given as follows:
\begin{eqnarray}
  Z[J] &=& \exp{i W [J]} = \langle \Omega_{\rm out} | \Omega_{\rm in} \rangle_J \\
  &=& \int DU DN DN^{\dagger} \langle \Omega_{\rm out} | N(+\infty) \rangle
  e^{i \int dx ({\cal L}_{\pi}+ {\cal L}_{\pi N}+ {\cal L}_{N})} \langle N(-\infty) | \Omega_{\rm in} \rangle ,
\end{eqnarray}
where $U$ is the chiral field parametrized by the NG boson field in the nonlinear realization of chiral symmetry, $N$ is the nucleon field, and $J$ represents the scalar, pseudo scalar, vector and axial vector external fields 
$J = (s,p,v,a)$. 
The Lagrangian is described by the free pion and nucleon fields and should
include, in principle, all the interaction among pions and nucleons within the Lagrangian. 
Since we use the Lagrangian described by the free nucleon field and the 
Lagrangian prescribes the nucleon interaction, 
the reference state can be the ground state of the free Fermi gas of nucleons defined by
\begin{equation}
  | \Omega_{\rm in,out} \rangle \equiv  \prod_n^N a^{\dag} ({\bf p}_n)| 0 \rangle ,
\label{generating}
\end{equation}
where $a^{\dag} ({\bf p}_n)$ is the nucleon creation operator with momentum $ {\bf p}_n $ and the index $n$ represents the spin and isospin and $N$ is 
the number of the momentum states below the nucleon Fermi momentum 
$k_F^{(n)}$, which is obtained by the nucleon density $\rho_{n}$ as 
$k_{F}^{(n)} = (3 \pi^2 \rho_n )^{1/3}$, and $ | 0 \rangle $ is the 0-particle state.

The connected $n$-point Green functions can be calculated by taking functional derivatives of $ i W [J] $ with respect to the external sources $J_i$:
\begin{equation}
\langle \Omega_{\rm out} |T {\cal O}_{1} \cdots {\cal O}_{n}  | \Omega_{\rm in} \rangle = (-i)^n \frac{\delta}{\delta J_{1} } \cdots \frac{\delta}{\delta J_{n} } i W[J] ,
\label{Green}
\end{equation}
where ${\cal O}_{i}$ is the corresponding current operator to the external source $J_{i}$.
Considering the symmetry property of the current operator under the chiral rotation,
one can identify the quark contents of the current operators,
such as the pseudo-scalar current $ P^i = \bar{q} i \gamma_5 \tau^i q $
and the axial-vector current $A^i_\mu = \bar{q} \gamma_\mu \gamma_5 \tau^i q/2 $.
If we evaluate the generating functional non-perturbatively with appropriate
nucleon-nucleon interactions, we can describe nuclear matter in principle 
and deduce the QCD current Green functions in nuclear matter.
Note that this prescription resembles the description of deuteron in terms of the free nucleons, in which one starts with the free nucleon field and with appropriate nucleon-nucleon interaction and 
one can find deuteron as a bound state of two nucleons by solving the Bethe-Salpeter equation non-perturbatively. 

Reference~\cite{Oll02} propose a method to derive the in-medium chiral Lagrangian
in a systematic expansion in terms of the chiral counting and the Fermi sea insertion.
In this method, the Fermi momentum is regarded as a small parameter as the 
chiral order. Performing the integral in terms of the nucleon field by 
using the Gauss integral formula for the bilinear form of the nucleon interaction,
one obtains the generating functional characterized by double expansion of 
Fermi sea insertions and chiral orders:
\begin{eqnarray}
\lefteqn{Z[J]=
 \int DU \exp \Big{\{} i\int dx \Big{[}\mathcal{L}_{\pi \pi} 
- \int \frac{d \bf{p}}{(2\pi)^3 2E_{p}} 
  {\mathcal F\,}
{\rm Tr} \Big{(} i\Gamma(x,y) (\slash{p} + m_N) n(p) \Big{)} } &&
\nonumber \\  && \hspace{-0.8cm}
- \frac{i}{2} \int \frac{d \bf{p}}{(2\pi)^3 2E_{p}} \frac{d \bf{q}}{(2\pi)^3 2E_{q}} 
  {\mathcal F}\,
 {\rm Tr} \Big{(} i\Gamma(x,x') (\slash{q} + m_N) 
 n(q) i\Gamma(y',y) (\slash{p} + m_N) n(p) \Big{)}
  + \cdots  ,
\Big{]} \Big{\}} \label{eq:Zexp}
\end{eqnarray}
where $\mathcal F$ denotes Fourier transformation of the spacial variables except for $x$, 
$ E({\bf p})$ is the nucleon energy $ E({\bf p}) = \sqrt{ {\bf p}^2 + m_N^2 }$ for 
the momentum $\bf p$ and $\Gamma(x,y)$ is the nonlocal vertex defined by
the in-vacuum quantities as $\Gamma \equiv -iA[1_4-G_0 A]^{-1}$, where 
the interactions $A$ and the free nucleon propagator $G_{0}$
are given by the  in-vacuum chiral perturbation theory. 
The isodoublet matrix $n(p)$ restricts the momentum integral for the nucleon 
momenta up to the Fermi momentum:
\begin{equation}
 n(p) = \left( 
       \begin{array}{cc}
        \theta ( k_F^{p} - | {\bf p} |) & 0  \\
           0        & \theta ( k_F^{n} - | {\bf p} |)   \\ 
       \end{array}         \right) .
\end{equation}
The nonlocal vertex $\Gamma(x,y)$ is expanded in terms of the bilinear local vertex $A$
as  
\begin{equation}
i \Gamma = A + A G_{0} A + A G_{0} A G_{0} A+ \cdots \label{eq:Gamma}
\end{equation}
and the vertex $A$ is also expanded in terms of the chiral order. 
The in-medium pion Lagrangian $ \tilde{\mathcal L}_{\pi \pi}$ can be obtained 
by the generating functional~\eqref{eq:Zexp}
as $Z[J]= \exp [{i \int d^{4}x  \tilde{\mathcal L}_{\pi \pi}}]$.

For the calculation of the 
amplitude, one uses the free nucleon propagators for the chiral expansion 
of the nonlocal vertex function while the Fermi sea nucleon term $ (- 2 \pi) \delta(p^2 - m_N^2) \theta (p_0) n(p)$ in the Fermi sea insertion among the nonlocal vertices $\Gamma(x,y)$.  
It has been shown explicitly in Ref.~\cite{God13} that
this expansion scheme is consistent with the conventional 
relativistic many body theory in the sense that one can use
directly the in-medium nucleon propagator, that is the Fermi gas propagator, 
\begin{eqnarray}
i G^i (p) &=& iG_0^i (p) + iG_m^i (p) \\
   &=& \frac{i(\slash{p} + m_N )}{p^2 - m_N^2 +i \epsilon } 
   - 2 \pi (\slash{p} + m_N) \delta (p^2 - m_N^2) \theta (p_0) \theta (k_F^i - |{\bf p}|)
\label{fermigas}
\end{eqnarray}
in the calculation. Here, the free propagator and the medium part of the Fermi gas propagator 
are denoted by $iG_0^i$, $iG_m^i $ with the isospin index $i = n,p$.
The medium part $iG_m^i $ represents Fermi sea effect.
We can calculate the connected functions using usual perturbative 
expansions with the in-medium nucleon propagator.
In this way, one can deal with density contributions from nuclear medium 
by perturbative expansion of nuclear Fermi momentum.

\subsection{Density expansion}
In the in-medium CHPT, we can classify the current Green functions in terms of
the order of the small parameters 
in the expansion of the pion momentum, the quark mass and 
the Fermi momentum in the similar way to the in-vacuum CHPT.
The chiral order of a specific diagram is counted as~\cite{Oll02}
\begin{eqnarray}
  \nu &=& 4 L_{\pi} - 2 I_{\pi} + \sum_{i=1}^{V_{\pi}} d_{i} + \sum_{i=1}^{V_{\rho}} d_{\rho i} \ge 4 \\
  d_{\rho} &=& 3n + \sum_{i=1}^{n} \nu_{\Gamma_{i}} - 4 (n-1)
\end{eqnarray}
where $L_{\pi}$ is the number of pion loops, $I_{\pi}$ is the number of the pion 
propagators, $d_{i}$ is the chiral dimension from the pion chiral Lagrangian,
$d_{\rho}$ is the chiral dimension of the nonlocal in-medium vertex with
$n$ Fermi sea insertions and $\nu_{\Gamma}$ is the chiral dimension of
the $\Gamma$ vertex. 
In this chiral counting, we count Fermi gas propagator as $O(p^{-1})$ like 
free nucleon propagator,
so that the counting rule is the same as in-vacuum chiral perturbation theory.

In this work, we focus on the Fermi momentum dependence of the Green 
functions and count only by Fermi momentum orders.
We assume that the in-vacuum loop effects are renormalized
into counter terms in the chiral Lagrangian
and we use the in-vacuum physical values to fix the low energy constants (LECs).
For example, we translate LEC $c_1$ appearing in $ \mathcal{L}_{\pi N}^{(2)}$ 
into the physical $\pi N $ sigma term $\sigma_{\pi N} $:
\begin{equation}
c_1 = - \frac{ \sigma_{\pi N}}{4 m_\pi^2},
\end{equation}
since the $c_{1}$ term gives the leading order of the $\pi N$ sigma term
in the chiral perturbation theory and once one calculates the higher orders
for the $\pi N$ sigma term, they should also contribute to the in-medium
quantities in the same manner as $c_{1}$.
An empirical value of the sigma term is $\sigma_{\pi N} \simeq 45 {\rm MeV} $ \cite{Gas91}.
and a recent analysis based on relativistic formulation of $\pi N$ chiral perturbation theory suggests $\sigma_{\pi N} = 59(7) {\rm MeV}$ \cite{Ala11,Ala12}.
Another example is the isoscalar $\pi N$ scattering length $a^+$.
In tree order, $a^+$ is given with LECs $c_1 , c_2 , c_3 $ from 
$\mathcal{L}_{\pi N}^{(2)} $ as
\begin{equation}
b^{+} \equiv {4 \pi} \Big{(} 1 + \frac{m_\pi}{m_N} \Big{)} a^+ =  \frac{2 m_\pi^2}{f_\pi^2} (2c_1 -c_2 -c_3 +\frac{g_A^2}{8m_N}) .
\end{equation}
The scattering length is used to determine the combination of the LECs $c_1 , c_2 , c_3 $.
Recent calculations based on CHPT give $ a^+ = (7.6 \pm 3.1 ) 10^{-3} m_\pi^{-1}$ at better than the 95$\%$ confidence level \cite{Bar11}.
In this way, we take the in-vacuum physical values to determine LECs 
and perform systematic calculations for density effects of the pionic observables
based on the counting of Fermi momentum orders.

As we have discussed in Ref.~\cite{God13}, beyond the order of $\rho^{2}$
in the density expansion, the $\pi N$ dynamics alone cannot predict 
the in-medium quantities, because we encounter divergence in loop 
calculations. These divergence can be removed, once we introduce counter 
terms expressed by $NN$ contact interactions. The $NN$ interactions
are to be determined by the in-vacuum $NN$ dynamics.

\subsection{Chiral Lagrangian and parametrization of chiral field}

We use the following chiral Lagrangian in this work. 
The chiral Lagrangian for the meson sector is given with the chiral field $U$ parametrized by the pion field as
\begin{equation}
   \mathcal{L}_{\pi}^{(2)} = 
    \frac{f^2}{4} {\rm Tr} \left( D_\mu U^\dag D^\mu U 
    + \chi^\dag U + \chi U^\dag \right) ,
\end{equation}
with the covariant derivative for the chiral field $U$ 
\begin{equation}
D_\mu U = \partial_{\mu} U -i ( v_{\mu} + a_{\mu} )U +iU( v_{\mu} - a_{\mu} ) 
\end{equation}
given by the vector and axial vector external fields $v_{\mu}$ and $a_{\mu}$
counted as $O(p)$, and $\chi$ field
\begin{equation}
\chi = 2B_0 (s + ip)
\end{equation}
given by the scalar and pseudoscalar fields $s$ and $p$ counted as $O(p^2)$.
The scalar field $s$ is replaced by the quark mass matrix in calculation.
This Lagrangian is counted as $O(p^2)$.

In this paper, in order to make the perturbative calculation simple, as we have 
done in the previous paper~\cite{God13}, we use the following parametrization 
of the chiral field $U$ proposed by \cite{Charap:1971bn,Gerstein:1971fm}:
\begin{equation}
   U = \exp\left[ i \pi^i \tau^i \frac{y(\pi^{2})}{2 \sqrt{\pi^{2}}} \right]
\end{equation}
where $y(\pi^{2})$ satisfies
\begin{equation}
   y - \sin y = \frac{4}{3} \left(\frac{\pi^{2}}{f^{2}} \right)^{\frac{3}{2}}.
\end{equation}

The chiral Lagrangian for the nucleon sector is as follows:
\begin{equation}
\mathcal{L}_{\pi N} =  \bar{N} (i \gamma^\mu \partial_\mu -m_N -A) N ,
\end{equation}
where $A$ is the chiral interaction for the nucleon in the bilinear form. 
The chiral interaction 
can be expanded in terms of the chiral order as
$
A = \sum_{n=1} A^{(n)}, 
$
and $ A^{(n)} $ is counted as $O(p^n)$. 
The explicit form of  the leading term $ A^{(1)} $ reads
\begin{equation}
   A^{(1)} = -i \gamma^{\mu} \Gamma_{\mu} 
     - i g_A \gamma^{\mu} \gamma_5 \Delta_{\mu} 
\end{equation}
with the vector current 
\begin{equation}
\Gamma_{\mu} = \frac{1}{2} [u^\dag , \partial_{\mu} u ] - \frac{i}{2} u^\dag ( v_{\mu} + a_{\mu} ) u - \frac{i}{2} u ( v_{\mu} - a_{\mu} ) u^\dag ,
\end{equation}
and the axial current
\begin{equation}
\Delta_{\mu} = \frac{1}{2} \Big{[} u^\dag \big{(} \partial_\mu - i (v_\mu + a_\mu ) \big{)} u  - u \big{(} \partial_\mu - i(v_\mu - a_\mu ) \big{)} u^\dag \Big{]},
\end{equation}
where the field $u$ is defined by a square root of the chiral field $U$: $ u = \sqrt{U} $. 
The expression of the next leading term $ A^{(2)} $ is given as
\begin{equation}
A^{(2)} = -c_1 \langle \chi_+ \rangle + \frac{c_2}{2m_N^2} \langle u_{\mu} u_{\nu} \rangle D^{\mu} D^{\nu} -\frac{c_3}{2} \langle u_{\mu} u^{\mu} \rangle
\end{equation}
with $\chi_+ \equiv u \chi^\dag u + u^\dag \chi u^\dag$, 
$ u_\mu \equiv 2i \Delta_\mu$ and 
the covariant derivative for the nucleon field
$ D_\mu N = (\partial_\mu  + \Gamma_\mu) N$.
Here we have omitted irrelevant terms in the present work for the in-medium 
pion properties in symmetry nuclear matter.

\section{In-medium properties of pion}
\label{decayconst}
\subsection{Pion mass and wave function renormalization}
The pion propagation in medium can be calculated by the two-point function of the pseudoscalar density:
\begin{equation}
  \Pi^{ab}(p) = \langle \Omega | P^a P^b  | \Omega \rangle .
\end{equation}
Around the in-medium pion pole the two-point function can be written in terms 
of the in-medium quantities, such as the in-medium pion mass $m_{\pi}^{*2}$, 
velocity $v_{\pi}$ and pseudoscalar coupling $G_\pi^*$ as
\begin{equation}
 \Pi^{ab}(p) =  \delta^{ab} G_\pi^* \frac{i}{p_{0}^2 - v_{\pi}^{2} {\bf p}^2 -  m_\pi^{*2} +i\epsilon} G_\pi^* \label{eq:pipropm}
\end{equation}
where $G_{\pi}^{*}$ does not include any singularity at the pion pole. 
In this way, with ${\bf p}=0$,
the in-medium pion mass is defined by the pole position of the two point function and
the coupling of pion to the pseudoscalar density $P^{a}$ in medium is defined by the
square root of the residue of the two-point function at the pion pole. 

The two-point function can be calculated using the in-medium chiral perturbation 
theory given in the previous section. The calculation will be done in terms of
the in-vacuum quantities like 
\begin{equation}
   \Pi^{ab}(p) =  \delta^{ab} \hat{G}_\pi \frac{i}{p^2 - m_\pi^2 - \Sigma (p^2) +i\epsilon} \hat{G}_\pi \label{eq:pipropv}
\end{equation}
where $m_{\pi}$ is the in-vacuum pion mass, $\Sigma(p^{2})$ is the pion self-energy
in medium and $\hat G_{\pi}$ is the vertex correction of the pion coupling to 
the pseudoscalar density. The vertex correction $\hat G_{\pi}$ does not contain 
the pion pole and it is calculated by considering one-particle irreducible diagrams. 

Expanding the self-energy $\Sigma(p^{2})$ in Eq.~\eqref{eq:pipropv} around 
$p_{0}^{2}=m_{\pi}^{*2}$ and ${\bf p}^2=0$,
$$
\Sigma(p^{2}) = \Sigma(m^{*2}_{\pi}) + (p_{0}^{2}-m_{\pi}^{*2}) \frac{\partial \Sigma(m_{\pi}^{*2})}{\partial p_{0}^{2}} 
+ {\bf p}^2\frac{\partial \Sigma(m_{\pi}^{*2})}{\partial {\bf p}^2} + \cdots,
$$
we write the two-point function in the following way
\begin{equation}
    \Pi^{ab}(p) =  \delta^{ab} \hat G_\pi \frac{iZ}{p_{0}^2 - v_{\pi}^{2} {\bf p}^2- m_\pi^{*2} +i\epsilon} \hat G_\pi \label{eq:pipropvm}
\end{equation}
with
\begin{eqnarray}
    m_{\pi}^{*2} &=& m_{\pi}^{2} + \Sigma(m_{\pi}^{*2}) \\
    v_{\pi}^{2} &=& 1 + \frac{\partial \Sigma(m_{\pi}^{*2})}{\partial {\bf p}^2} \\
    Z &=&  \Big{(} 1 - \frac{\partial \Sigma}{\partial p_0^2} (m_{\pi}^{*2}) \Big{)}^{-1} 
\end{eqnarray}
Comparing Eqs.~\eqref{eq:pipropm} and \eqref{eq:pipropvm}, we obtain, at the pion pole, 
\begin{equation}
    G_{\pi}^{*} = \sqrt Z \hat G_{\pi} \label{eq:hatmed} .
\end{equation}

Using the in-medium chiral perturbation theory we can calculate the self-energy 
$\Sigma(p^{2})$ and $\hat G_{\pi}$, we obtain the in-medium pion properties
with the above equations. 

\subsection{In-medium state} 

The in-medium pion propagator can be also written by the pion field operator 
$\pi$ as
\begin{eqnarray}
   \langle \Omega | \pi^{a} \pi^{b} | \Omega \rangle 
  &=& \delta^{ab} \frac{i}{p^{2} - m_{\pi}^{2} - \Sigma(p^{2}) + i\epsilon} \\
  & =& \delta^{ab} \frac{iZ}{p^{2}_{0} - v_{\pi}^{2} {\bf p}^2 - m_{\pi}^{*2} + i\epsilon} \label{eq:pipiv}.
\end{eqnarray}
Comparing Eqs.~\eqref{eq:pipropvm} and \eqref{eq:pipiv}, for the calculation of the pion pole 
we regard 
\begin{equation}
\pi^{a}= \frac{P^{a}}{\hat G_{\pi}} \label{eq:Phatpi} .
\end{equation}
The pion operator $\pi$ creates one pion in medium with the mass $m_{\pi}^{*}$
and the wave function normalization $Z$, satisfying
\begin{equation}
  \langle \Omega | \pi^{a} | \pi^{*b}(p) \rangle = \delta^{ab} \sqrt{Z} ,
\end{equation}
where we have introduced the one-pion state with a momentum $p$ 
in the nuclear medium by denoting $| \pi^{*b}\rangle$.

The in-medium coupling constant $G_{\pi}^{*}$ defined in Eq.~\eqref{eq:pipropm}
as the residue of the pion propagator induced by the pseudoscalar density
may be also written as the following matrix element:
\begin{equation}
   G_{\pi}^{*} \delta^{ab} = \langle \Omega | P^{a} | \pi^{*b}(p) \rangle \label{eq:Gpistdef}
\end{equation}

Relation~\eqref{eq:hatmed} can be understood by Eq.~\eqref{eq:Gpistdef} with 
the reduction formula:
\begin{eqnarray*}
   \lefteqn{
   \langle \Omega | P^{a} | \pi^{*b}(p) \rangle 
   } && \nonumber \\ 
   &=& \lim_{p^{2} \to m_{\pi}^{*2}}
   \left(\frac{i\sqrt Z}{p^{2} - m_{\pi}^{*2} + i\epsilon}\right)^{-1}
   \langle \Omega | P^{a} \pi^{b} | \Omega \rangle \\
   &=& \lim_{p^{2} \to m_{\pi}^{*2}}
   \left(\frac{i \sqrt Z}{p^{2} - m_{\pi}^{*2} + i\epsilon}\right)^{-1} \frac{1}{\hat G_{\pi}}
   \langle \Omega | P^{a} P^{b} | \Omega \rangle \\ 
   &=& \lim_{p^{2} \to m_{\pi}^{*2}}
   \left(\frac{i \sqrt Z}{p^{2} - m_{\pi}^{*2} + i\epsilon}\right)^{-1} \frac{1}{\hat G_{\pi}}
  \delta^{ab} \left( \frac{\hat G_{\pi} i Z \hat G_{\pi}}{p^{2} - m_{\pi}^{2} + i\epsilon} \right)\\ 
   &=& \delta^{ab} \sqrt Z \hat G_{\pi} 
\end{eqnarray*}
where we have understood $p^{2} = p_{0}^{2} - v_{\pi}^{2} {\bf p}^2$,
and we have used Eq.~\eqref{eq:Phatpi}
in the second equality and Eq.~\eqref{eq:pipropvm} in the third equality.

\subsection{In-medium pion decay constants} 

We define the in-medium decay constant in analogy of the in-vacuum decay constant
as a matrix element of the axial vector current $A_{\mu}^{a}$:
\begin{equation}
\langle \Omega | A_\mu^a (0) | \pi^{*b} (p) \rangle = i \big{[} p_\mu F^* (p_{0},{\bf p})  
+ n_\mu (p \cdot n) N^* (p_{0}, {\bf p}) \big{]} \delta^{ab} \label{eq:pidecaycon}
\end{equation}
where we have introduced a vector characterizing the medium rest frame $ n_\mu $ 
with $n^{2}=1$ and there are two form factors, 
$F^*$ and $N^*$, in the presence of nuclear matter.  These form factors, actually, 
should be functions of $p^{2}$ and $p\cdot n$ according to Lorentz covariance. 
The pion decay constants are obtained at the mass shell point with $|{\bf p}| = 0$ and
$p_{0} = m_{\pi}^{*}$. 
We define the temporal and spatial components of the decay constant $f_t , f_s$
by taking $n_{\mu}= (1, {\bf 0})$ as
\begin{eqnarray}
   \langle \Omega| A_{0}^{a} | \pi^{*b} \rangle & \equiv & i f_{t} p_{0} , \\
   \langle \Omega| A_{i}^{a} | \pi^{*b} \rangle & \equiv & i f_{s} p_{i} .
\end{eqnarray}
The decay constants are obtained by
\begin{eqnarray}
f_t & = & F^*(m_{\pi}^{*},\vec 0) +N^*(m_{\pi}^{*},\vec 0) ,\\
f_s & = & N^* (m_{\pi}^{*},\vec 0) .
\end{eqnarray}

These decay constants can be calculated in the in-medium chiral perturbation 
theory. 
Making good use of the reduction formula in momentum space again, we 
write down the matrix element in terms of the one-particle irreducible 
vertex correction and the wave function renormalization:
\begin{eqnarray*}
   \lefteqn{
      \langle \Omega | A_\mu^a | \pi^{*b} (p) \rangle  
    } && \nonumber \\
 &=& \lim_{p^2 \to m_\pi^{*2}} \Big{(} \frac{i \sqrt{Z} }{p^2 - m_\pi^{*2} +i\epsilon} \Big{)}^{-1} \langle \Omega | A_\mu^a \pi^b | \Omega \rangle \\
 &=& \lim_{p^2 \to m_\pi^{*2}} \Big{(} \frac{i \sqrt{Z} }{p^2 - m_\pi^{*2} +i\epsilon} \Big{)}^{-1} \frac{ 1 }{ \hat{G}_\pi } \langle \Omega | A_\mu^a P^b | \Omega \rangle \nonumber \\
&=& \lim_{p^2 \to m_\pi^{*2}} \Big{(} \frac{i \sqrt{Z} }{p^2 - m_\pi^{*2} +i\epsilon} \Big{)}^{-1} 
\frac{ 1 }{ \hat{G}_\pi } i \hat{f}_i p_\mu \Big{(} \frac{iZ}{p^2 - m_\pi^{*2} +i\epsilon} \Big{)} \hat{G}_\pi \nonumber \\
&=& i \hat{f}_i \sqrt{Z}  p_\mu.
\end{eqnarray*}
where we mean $p^{2} = p_{0}^{2} - v_{\pi}^{2} {\bf p}^{\, 2}$
and $\hat{f}_i$ with $i = t,s$ is the vertex correction of the decay constant, which 
can be calculated by one-particle irreducible diagrams in the chiral perturbation theory.

\section{Results}
\label{results}
In this section, we investigate explicitly the in-medium pion properties
with the in-medium chiral perturbation theory up to the order of $k_{F}^{4}$
in the density expansion. 
In last section, we find that the in-medium pion decay constant 
can be calculated by the pion wave function renormalization $Z$
and the one-particle irreducible vertex correction $ \hat{f}_i $.
The wave function renormalization is evaluated by
taking derivative of the pion self-energy $\Sigma$ with respect to the energy squared.
We calculate the pion self energy first.  With the self energy, we evaluate
next the in-medium pion mass and the wave function renormalization,
and show their density dependence. 
Calculating the one-particle irreducible correction for the pion decay constant,
we evaluate the density dependence of the decay constant with  
the wave function renormalization.
We also check whether the low energy 
relations are satisfied also in the nuclear medium. Finally we discuss 
the $\pi^{0} \to 2\gamma$ decay rate in nuclear medium. 
For simplicity, we concentrate on the calculation 
under the unpolarized symmetric nuclear matter, where the matter has spin 0 
and isospin 0 with the equal proton and neutron densities, $\rho_{p}=\rho_{n}$.

In the following, we write down the Feynman diagrams for each
quantity and classify them according to the density dependence. 
As we have already discussed in the previous section, 
we presume that the in-vacuum LECs are fixed by the experimental 
observables. The vertex and mass corrections from the quantum 
loops of pion and nucleon are renormalized into the physical quantities. 
On this understanding, we replace LECs to the 
observed quantities and use the experimental values. 
In the present work the relevant replacements
are the followings:
\begin{eqnarray}
   2 B_{0} m_{q} &\to & m_{\pi}^{2}  \label{eq:B0} \\
   c_{1} & \to & - \frac{\sigma_{\pi N}}{4m_{\pi}^{2}} \\
  2c_{1} - c_{2} - c_{3} &\to & \frac{f_{\pi}^{2}}{2m_{\pi}^{2}} b^{+} - \frac{g_{A}^{2}}{8m_{N}} \label{eq:c1c2c3}
\end{eqnarray}
where the left hand sides are given by LECs, while the right hand sides
are written in terms of the observed quantities in vacuum. In these expression,
$b^{+} \equiv {4 \pi} ( 1 + m_\pi / m_N ) a^+ $ with the
isoscalar $\pi N$ scattering length $a^{+}$.
In the replacement
higher order contributions in the chiral counting are already involved.
In this sense, we loose strict counting of the chiral order.
Taking this scheme, we do not have to calculate the loop integrals which contain
only the in-vacuum propagators, because they are supposed to be already 
counted inside the experimental value. This fact reduces the number of the 
relevant diagrams which we should calculate. 

The in-medium quantities which we are going to calculate should be 
evaluated at the pion pole. The in-vacuum chiral order can be counted by the pion mass.
Therefore, in the perturbative expansion, the term of which the leading
chiral counting in vacuum is $\ell$ has 
$m_\pi^\ell k_F^m ( \frac{k_F}{m_N})^n $ dependence. 
The factor $ (k_F / m_N)^n $ comes from the Fermi motion correction of the 
nucleon when one calculates the Fermi sea loop integral.
The order of the density expansion is counted as $p= m+n$, while
the order of the small parameter is given by $q=\ell + m$.
If one takes the density contribution up to $k_{F}^{4}\sim \rho^{4/3}$, 
it is enough to consider the diagrams with $q \le m+4$. 

The details of the loop calculation is summarized in appendix~\ref{app:calc}.

\subsection{In-medium pion self energy}
\label{mass}

\begin{figure}[t]
  \begin{center}
  \includegraphics[ width=10cm ]{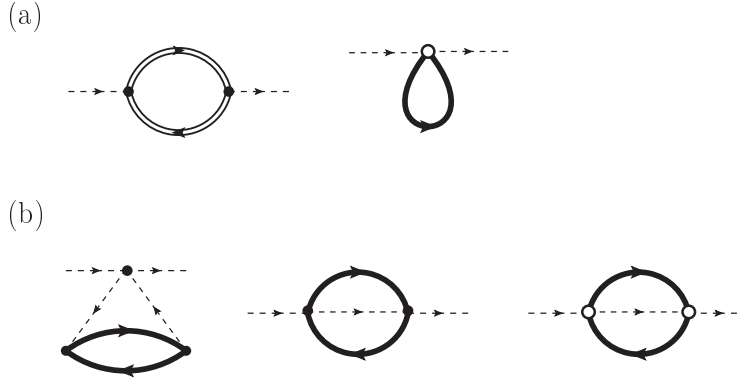}
  \caption{
  Feynman diagrams for the pion self energy $\Sigma$ up to $O(k_{F}^{4})$.
 (a) the leading order of the density corrections $O(k_{F}^{3})$. 
 (b) the next-to-leading order correction $O(k_{F}^{4})$.
In these diagrams, the dashed lines denote the pion lines,  the doubled and thick lines are
the nucleon lines for Fermi gas propagator $iG$ and the medium part of the Fermi gas propagator $i G_m$, 
the filled and unfilled circles are the leading and the next-to-leading order vertices  
from the chiral Lagrangian respectively.}
  \label{self}
  \end{center}
\end{figure}

We show the Feynman diagrams contributing the in-medium self energy with the density corrections up to $ O(k_F^4  \sim \rho^{4/3}) $ in 
Fig. \ref{self}.
In these diagrams, the dashed doubled and thick lines are pion propagator, the Fermi gas propagator $iG$ and the nucleon propagation in the Fermi gas $iG_m$,
and the filled and unfilled circles are the leading and next-to-leading order vertices 
from the chiral Lagrangian, respectively.
In the following calculations, we fix the external momentum as $ q= (q_0, {\bf 0}) $.

The diagrams for the leading order of the density expansion $O(k_{F}^{3})$ are given in Fig.~\ref{self}(a). 
The self energy coming from the left diagram $\Sigma_1$ in Fig.\ref{self} (a) can be calculated as
\begin{eqnarray}
-i \Sigma_1 (q_0) &=& - \int \frac{d^4 p}{(2 \pi )^4} {\rm Tr} \Big{[} (-i A^{(1)}_\pi) i G(p+q) (-i A^{(1)}_\pi ) iG(p) \Big{]} \nonumber \\
&=& - \rho \frac{i g_A^2}{4 f^2 m_N}  q_0^2,
\end{eqnarray}
where $iG(p)$ is the Fermi gas propagator given in Eq.~\eqref{fermigas},
the amplitude $-i A_{\pi}^{(1)}$ can be obtained by the interaction 
Lagrangian given in Eq.~\eqref{eq:piNN} of appendix~\ref{app:lagrangian}.

Next, we calculate the right diagram $\Sigma_2$ in Fig.\ref{self}(a):
\begin{eqnarray}
-i \Sigma_2 (q_0) &=& (-1) \int \frac{d^4 p}{(2\pi)^4} {\rm Tr} \Big{[} (-iA_{\pi \pi}^{(2)}) iG_m (p) \Big{]} \nonumber \\
&=& -\frac{2 i\rho}{f^2} \Big{(} 4 c_1 B_{0} m_{q} - ( c_2 + c_3 ) q_0^2  \Big{)} .
\end{eqnarray}

The next-to-leading order contribution coming from the left diagram $\Sigma_3$ in Fig.~\ref{self}(b)
is given as
\begin{equation}
-i\Sigma_3 (q_0) = \frac{1}{2} \int \frac{d^4 k}{(2\pi)^4} i {\mathcal L}_{\pi^4}^{(2)}  (i D_\pi (k) )^2 (-i\Sigma_m (k)) 
\end{equation}
with the pion propagator $iD_{\pi}(k)$, the symmetric factor $1/2$ and 
$\Sigma_m (k)$ defined as
\begin{displaymath}
-i \Sigma_m (k) = - \int \frac{d^4 p}{(2 \pi )^4} {\rm Tr} \Big{[} (-i A^{(1)}_\pi) i G_m(p+k) (-i A^{(1)}_\pi ) iG_m(p) \Big{]} .
\end{displaymath}
Finally we obtain 
\begin{eqnarray}
- i \Sigma_3 (q_0)
&=& \frac{i m_\pi^2 g_A^2 k_F^4 }{12 \pi^4 f^4} F(\frac{m_{\pi}^{2}}{4k_{F}^{2}})
\end{eqnarray}
with function $F(x^2)$ defined by
\begin{equation}
  F(x^{2}) =  \frac{3}{8} -\frac{3x^2}{4} -\frac{3x}{2} {\rm arctan} \frac{1}{x} + \frac{3x^2}{4} (x^2 +2) \ln \frac{x^2 +1}{x^2} 
\label{functionf}
\end{equation}

The self energies coming from the double scattering corrections \cite{Eri66} given by the middle and right diagrams of Fig.~\ref{self}(b), $\Sigma_{4}$ and $\Sigma_{5}$, have the two $\pi\pi NN$ vertices from the leading Weinberg-Tomozawa interaction and the next-to-leading interaction with LECs $c_{i}$, respectively. 
These are evaluated as
\begin{eqnarray}
-i\Sigma_4 (q_0) &=& - \int \frac{d^4 p}{(2 \pi )^4} \frac{d^4 k}{(2 \pi )^4} {\rm Tr} \Big{[} (-i A^{(1)}_{\pi \pi}) i G_m (k- \frac{p}{2} ) (-i A^{(1)}_{\pi \pi} ) iG_m (k+ \frac{p}{2}) iD_\pi (p+q) \Big{]} \nonumber \\
&=&  -\frac{2 i q_0^2 }{f^4} \frac{k_F^4}{6 \pi^4} G(a^{2}) ,\\
-i\Sigma_5 (q_0) &=& - \int \frac{d^4 p}{(2 \pi )^4} \frac{d^4 k}{(2 \pi )^4} {\rm Tr} \Big{[} (-i A^{(2)}_{\pi \pi}) i G_m (k- \frac{p}{2} ) (-i A^{(2)}_{\pi \pi} ) iG_m (k+ \frac{p}{2}) iD_\pi (q-p) \Big{]} \nonumber \\
&=& -i \Big{(} 8c_1 B_0 m_q  - 2c_2 q_0^2 - 2c_3 q_0^2 \Big{)}^2 \frac{2 k_F^4}{3 \pi^4 f^4} G(a^{2}) 
\end{eqnarray}
with $a^{2} = (q_{0}^{2} - m_{\pi}^{2})/(4k_{F}^{2})$ and function $G(a^{2})$ defined by 
\begin{equation}
  G(x^{2}) = \frac{3}{8} + \frac{x^{2}}{4}  + \frac{\sqrt{x^{2}}}{2} \ln | \frac{1- x}{1+ x} | + \frac{x^{2}}{4} (x^{2} -3) \ln | \frac{1-x^{2}}{x^{2}} | .
\label{functiong}
\end{equation}

Summing up all the contributions, we obtain the the self energy up to $O(k_{F}^{4})$ as
\begin{eqnarray}
\Sigma (q_0 , {\bf 0}) &=& \Sigma_1 + \Sigma_2 + \Sigma_3 + \Sigma_4 + \Sigma_5  \\
&=&  \frac{2 \rho}{f^2} \Big{(} 4 c_1 B_{0} m_{q} - ( c_2 + c_3 - \frac{g_A^2}{8 m_N} ) q_0^2  \Big{)} - \frac{ m_\pi^2 g_A^2 k_F^4 }{12 \pi^4 f^4}  F(\frac{m_\pi}{2 k_F}) \nonumber \\
&& + \left[\frac{q_{0}^{2}}{8} + \Big{(} 4c_1 B_0 m_q  - c_2 q_0^2 - c_3 q_0^2 \Big{)}^2 \right] \frac{8k_F^4}{3 f^4 \pi^4} G(a^{2}) 
\end{eqnarray}
with $a^{2} = (q_{0}^{2} - m_{\pi}^{2})/(4k_{F}^{2})$.

\subsection{In-medium pion mass}
The in-medium pion mass is obtained by the summation of the in-vacuum mass and 
the self energy evaluated at the pion on the mass shell $q^{\mu}=(m_{\pi}^{*},{\bf 0})$.
This brings us a self-consistent equation:
\begin{equation}
m_\pi^{*2} \equiv  m_\pi^2 + \Sigma (q_0^2= m_\pi^{*2}) .
\end{equation}
Nevertheless, because the in-medium correction starts with the linear density 
$\rho \sim k_{F}^{3}$ and we are evaluating the pion mass up to $O(\rho^{4/3})$,
the density correction on the mass in the argument of the self energy gives
higher orders in the density expansion of the in-medium pion mass. Thus,
we are allowed to evaluate the self-energy at the in-vacuum on-shell
$q^{\mu}=(m_{\pi},{\bf 0})$ for the present purpose. We evaluate the 
in-medium pion mass up to $O(\rho^{4/3})$ as
\begin{eqnarray}
m_\pi &= & \sqrt{m_\pi^2 + \Sigma (q_0^2= m_\pi^{2})} \nonumber \\
&=& m_\pi \left\{ 1 + \frac{\rho}{f^2}( 2 c_1 -c_2 - c_3 + \frac{g_A^2}{8 m_N} ) - \frac{ g_A^2 k_F^4 }{24 \pi^4 f^4} F(\frac{m_\pi}{2k_F}) 
\right. \nonumber \\ && \left.
 + \left[ \frac{1}{8} + m_\pi^2 \Big{(} 2c_1 - c_2 - c_3 \Big{)}^2\right] \frac{4 k_F^4}{3 \pi^4 f^4} G(0) \right\} \nonumber \\
&=& m_\pi \left\{ 1 + \rho \frac{b^{+}}{2m_\pi^2} - \frac{ g_A^2 k_F^4 }{24 \pi^4 f_\pi^4} F(\frac{m_\pi}{2k_F}) 
 + \left[ \frac{1}{8} + m_\pi^2 \Big{(} \frac{b^{+}}{2m_\pi^2}-\frac{g_A^2}{8m_N} \Big{)}^2 \right]\frac{ 2k_F^4}{\pi^4 f_\pi^4}  \right\} .
\end{eqnarray}
The LECs, $B_{0}$, $c_{1}$, $c_{2}$, $c_{3}$, have been determined 
by the in-vacuum physical quantities, the pion mass $m_{\pi}$, 
the $\pi N$ sigma term $\sigma_{\pi N}$ and factored scattering length $b^{+}$. 
The linear density correction of the pion mass stems from the scattering length.
Since the isosinglet $\pi N$ scattering length is known to be a small number 
compared to the inverse pion mass, $ b^+ = (9.6 \pm 3.9) 10^{-2} m_\pi^{-1}$,
the leading correction is as small as 5\% at the saturation density 
$\rho_{0} \simeq 0.49\, m_{\pi}^{3}$. 

\begin{figure}[t]
    \centering
    \includegraphics[width=6cm,angle=-90]{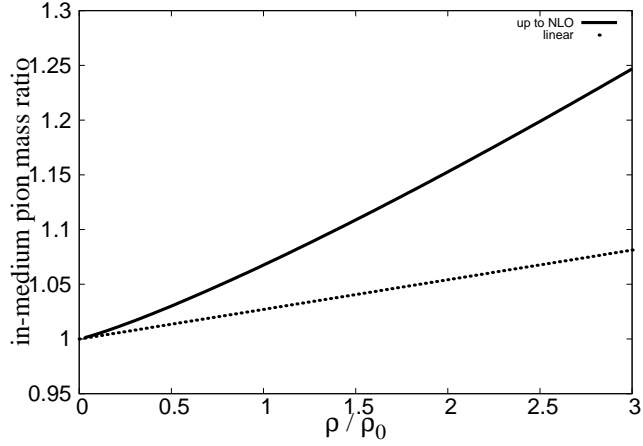}
    \caption{Density dependence of the in-medium pion mass normalized by the in-vacuum 
    pion mass, $m_\pi^{*} / m_\pi$, in symmetric nuclear matter.
    The dotted line shows the result up to the leading linear density correction, 
    while the solid line is the result containing the next-to-leading order correction.}
    \label{mass_dep}
  \end{figure}

In Fig.~\ref{mass_dep}, we show
the density dependence of the in-medium pion mass as a function of 
the density normalized by the normal nuclear density $\rho_0$. In the figure,
the dotted line shows the result up to the leading linear density and the solid line is 
for the result containing the next-to-leading order.
We take the following values of the in-vacuum quantities; $ f_\pi = 92.4{\rm MeV}$, $g_A =1.26$, $m_\pi = 138 {\rm MeV}$, $b^{+}= 9.6 \times 10^{-2} m_\pi^{-1}$ and $\sigma_{\pi N} = 45 {\rm MeV}$~\cite{Gas91}.
One can see from Fig.~\ref{mass_dep} that the NLO correction is not small. 
The main contribution comes from the double scattering terms. 
The density correction of the pion mass at twice or three times the normal 
nuclear density becomes about 15 to 20\%. 
Since the in-medium CHPT which is a low energy effective theory 
it would be not applicable in higher density region,
nevertheless we expect that this theory would be applicable 
up to 3$\rho_0$ where Fermi momentum corresponds to about 400 MeV.

\subsection{In-medium wave function renormalization}
Next, the wave function renormalization is obtained by evaluating the derivative of 
the self energy $\Sigma (q_0)$ with respect to $q_0^2$ at $q_{0}^{2} = m_{\pi}^{*2}$.
Again since the difference between the in-medium and in-vacuum pion masses in
the self energy is counted as the higher order of the density expansion, 
we evaluate the derivative of the self energy at $q_{0}^{2} = m_{\pi}^{2}$ for the
present purpose:
\begin{eqnarray}
Z &= & \Big{[} 1- \frac{ \partial \Sigma (q_0^2 = m_\pi^{2} )}{\partial q_0^2 } \Big{]}^{-1} \nonumber \\
&=& \left[ 1 + \frac{2\rho}{f^2}( c_2 + c_3 - \frac{ g_A^2}{8 m_N} ) 
- \left(\frac{1}{8m_\pi^2} - 2  ( 2c_1 - c_2 - c_3 ) (c_2 +c_3)\right) 
\frac{8 m_\pi^2 k_F^4}{3 \pi^4 f^4}G(0) \right]^{-1} \nonumber \\
&=& 1 + \frac{2\rho}{f_\pi^2} \Big{(} \frac{\sigma_{\pi N} }{2 m_\pi^2} 
+ \frac{ f_\pi^2 b^{+}}{2 m_\pi^2 } \Big{)} + \frac{m_\pi^2 k_F^4}{\pi^4 f_\pi^4} \left[ \frac{1}{8m_\pi^2} +
 2\Big{(} \frac{ f_\pi^2b^{+}}{2m_\pi^2 } - \frac{g_A^2}{8m_N} \Big{)}
\Big{(} \frac{\sigma_{\pi N}}{2 m_\pi^2} +  \frac{f_\pi^2 b^{+}}{2m_\pi^2 } - \frac{g_A^2}{8m_N }  \Big{)} \right] . \nonumber \\
\end{eqnarray}
In the calculation of the wave function renormalization, we encounter a singularity 
in the derivative of function $G(x)$ at $x\to 0$. As discussed in detail in 
appendix~\ref{infra}, this singularity may be considered as infrared singularity in 
quantum field theory and brings an problematic density dependence starting 
from $k_{F}^{2} \ln k_{F}$, which is inconsistent with the low density expansion
starting from $k_{F}^{3}$. 
Thus, we should expect certain cancellation of the singularity. 
The infrared singularity should cancel when one calculates scattering rates, not 
in the amplitude itself. Therefore we have dropped the singular term
containing the derivative of function $G(x)$.

  \begin{figure}[t]
    \centering
    \includegraphics[width=6cm,angle=-90]{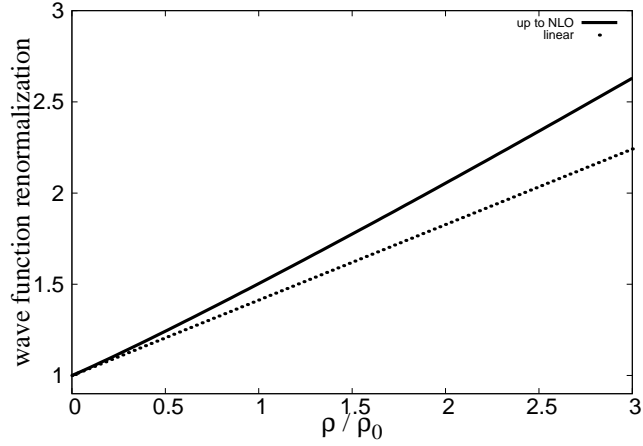}
    \caption{Density dependence of the in-medium pion wave function renormalization $Z$ in symmetric nuclear matter.
    The dotted line shows the result up to the leading linear density correction and the solid line is for the pion mass containing the next-to-leading order correction.}
    \label{wave}
  \end{figure}

At the linear density, we find that 
\begin{equation}
   Z = 1 + 0.40 \frac{\rho}{\rho_0}
\end{equation}
which implies that the wave function is enhanced by 40\% at the saturation density. 
The density dependence of wave function renormalization is shown in Fig.\ref{wave}.
As one can see in the figure, the density correction of the wave function is not 
so small, being an order of 50\% at the saturation density. It would be very interesting, 
if one could observe the strong enhancement of the wave function renormalization 
phenomenologically, such as formation cross sections of deeply bound pionic atoms.
Later, we will discuss again possible observation of the wave function renormalization  
itself in the in-medium $\pi^{0}$ decay into $2\gamma$.

\subsection{In-medium pion decay constant}
\label{constant}
We evaluate the temporal pion decay constant in medium.
According to the discussion in the previous section, the in-medium decay
constant is given by the wave function renormalization $Z$ and the
one-particle irreducible vertex correction $\hat f_{t}$ as $f_{t} = \hat f_{t} \sqrt Z$.
The wave function renormalization has been already evaluated 
in the previous subsection, here let us calculate the vertex correction 
$\hat{f}_t$ in density expansion.

\begin{figure}[t]
  \begin{center}
  \includegraphics[ width=10cm ]{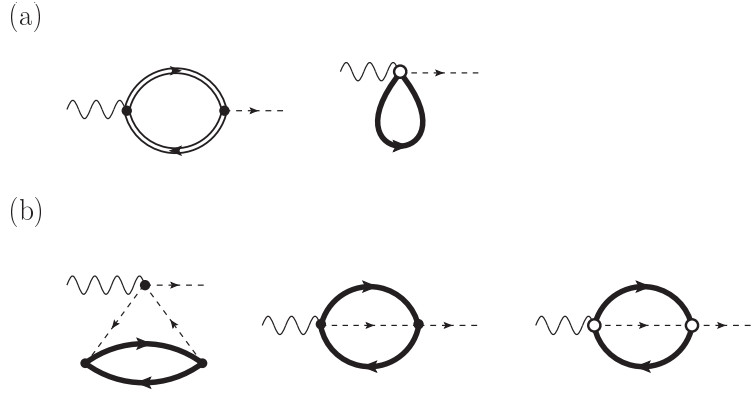}
  \caption{Feynman diagrams contributing the one-particle vertex correction 
  of the pion decay constant $\hat{f}_t$. (a) leading order density corrections.
  (b) the next-to-leading order corrections. 
 In these diagrams, the wavy line denotes axial vector current.}
  \label{decaygraph}
  \end{center}
\end{figure}

In Fig.~\ref{decaygraph}, we show the relevant Feynman diagrams 
for the density corrections up to $ O(k_F^4  \sim \rho^{4/3}) $.
The wavy lines in the Fig.\ref{decaygraph} denote axial vector currents.
The leading graphs $O(k_{F}^{3})$ are shown in Fig.~\ref{decaygraph}(a), while 
the next-to-leading order contributions $O(k_{F}^{4})$ are 
given in Fig.~\ref{decaygraph}(b).
Since the temporal decay constant is given by
the matrix element of the time-component of the axial vector current $A_{0}^i$,
we evaluate the one-particle irreducible (1PI) matrix element 
\begin{equation}
   \langle \Omega | A_{0}^{i} | \pi^{*j}(q) \rangle_{\rm 1PI} 
   = i \delta^{ij} p_{0} \hat f_{t}
\end{equation}
for the rest pion  $q_\mu = (q_{0}, {\bf 0} ) $.
The loop calculation is essentially same as the calculation done in the 
self-energy. Here we will show only the definition and result of each contribution.

The contribution from the left diagram $\hat{f}_1 $ in Fig.~\ref{decaygraph}(a)
reads
\begin{equation}
\hat{f}_1 q_0 
= - \int \frac{d^4 p}{(2 \pi )^4} {\rm Tr} \Big{[} (-i A^{(1)}_a) i G(p+q) (-i A^{(1)}_\pi ) iG(p) \Big{]}  = -  q_0\frac{ g_A^2}{4 f m_N} \rho  .
\end{equation}
The vertex correction coming from the right diagram in Fig.~\ref{decaygraph}(a) is
\begin{equation}
\hat{f}_2 q_0
= - \int \frac{d^4 p}{(2\pi)^4} {\rm Tr} \Big{[} (-iA_{\pi a}^{(2)}) i G_m (p) \Big{]} 
= q_0 \frac{2 \rho}{f} (c_2 + c_3) .
\end{equation}
For the left diagram in Fig.~\ref{decaygraph}(b), we have $\hat{f}_3 $ with symmetric factor $1/2$:
\begin{eqnarray}
\hat{f}_3 q_0
&=& \frac{1}{2} \int \frac{d^4 k}{(2\pi)^4} i {\mathcal L}_{\pi^3 a}^{(2)} \big{(} i D_\pi (k) \big{)}^2 (-i \Sigma_m ) \nonumber \\
&=& q_0  \frac{g_A^2 k_F^4}{6 \pi^4 f^3} F(\frac{m_\pi^2}{4 k_F^2}).
\end{eqnarray}
Function $F(x)$ is defined in Eq.~\eqref{functionf}.
The vertex corrections represented as the middle and right diagrams 
in Fig.~\ref{decaygraph}(b) are calculated as
\begin{eqnarray}
\hat{f}_4 q_0 &=& - \int \frac{d^4 p}{(2 \pi )^4} \frac{d^4 k}{(2 \pi )^4} {\rm Tr} \Big{[} (-i A^{(1)}_{\pi a}) i G_m (k- \frac{p}{2} ) (-i A^{(1)}_{\pi \pi} ) iG_m (k+ \frac{p}{2}) iD_\pi (p+q) \Big{]}\nonumber \\
& = &- \frac{q_0 k_F^4}{3 \pi^4 f^4} G(a^{2}) \\
  \hat{f}_5 q_0 &=&  - \int \frac{d^4 p}{(2 \pi )^4} \frac{d^4 k}{(2 \pi )^4} {\rm Tr} \Big{[} (-i A^{(2)}_{\pi a}) i G_m (k- \frac{p}{2} ) (-i A^{(2)}_{\pi \pi} ) iG_m (k+ \frac{p}{2}) iD_\pi (q-p) \Big{]} \nonumber \\
  &=&  (c_2+ c_3) \Big{(} 2c_1m_\pi^2 - (c_2+ c_3)q_0^2 \Big{)} \frac{8 q_0 k_F^4}{3 \pi^4 f^3} G(a^{2})
\end{eqnarray}
where $a^{2} = (q_{0}^{2}-m_{\pi}^{2})/(4k_{F}^{2})$ and function $G(x)$ is 
defined in Eq.~\eqref{functiong}.

The vertex correction $\hat f_{t}$ should be evaluated at $q_{0}^2 =m_{\pi}^{*2}$ in 
principle, because the decay constant is the matrix element of the axial 
current of the in-medium pion state having the on-shell condition $q_{0}=m_{\pi}^{*}$.
But, in practice, we are allowed to evaluate the vertex correction 
at $q_{0}=m_{\pi}$, because the difference is in the higher density orders. 

Summing up all the contributions, 
we obtain the density dependence of $ \hat{f}_t $ up to $ O ( k_F^4) $:
\begin{eqnarray}
\hat{f}_t (q_0 = m_\pi) &=& \hat{f}_1 +\hat{f}_2 +\hat{f}_3 +\hat{f}_4 +\hat{f}_5 \\
&=& f  + \frac{2 \rho}{f} (c_2 + c_3-\frac{g_{A}^{2}}{8m_{N}}) + \frac{g_A^2 k_F^4}{6 \pi^4 f^3} F(\frac{m_\pi}{2k_F}) 
 \nonumber \\ && 
 - \left[\frac{1}{8} - (c_2+ c_3) ( 2c_1 - c_2 - c_3 ) \right]\frac{8 m_\pi^2 k_F^4}{3 \pi^4 f^3} G(0) \nonumber \\
&=& f_\pi \Big{[}  1 - \frac{2 \rho}{f_\pi^2} \Big{(} \frac{\sigma_{\pi N} }{2 m_\pi^2} 
+ \frac{ f_\pi^2 b^{+}}{2 m_\pi^2 } \Big{)} + \frac{g_A^2 k_F^4}{6 \pi^4 f_\pi^4}  F(\frac{m_\pi}{2k_F}) 
\nonumber \\ && 
  - \frac{m_\pi^2  k_F^4}{\pi^4 f_\pi^4} 
    \left[\frac{1}{8} + \Big{(} \frac{ f_\pi^2b^{+}}{2m_\pi^2 } - \frac{g_A^2}{8m_N} \Big{)}
\Big{(} \frac{\sigma_{\pi N}}{2 m_\pi^2} +  \frac{f_\pi^2 b^{+}}{2m_\pi^2 } - \frac{g_A^2}{8m_N }  \Big{)} \right] 
\end{eqnarray}
where we have used $G(0)=3/8$.

With $\hat f_{t}$, we can calculate in-medium pion decay constant:
\begin{eqnarray}
f_t &=& \hat{f} \sqrt{Z} \nonumber \\
&=& f_\pi \Big{[} 1 - \frac{\rho }{f_\pi^2} \Big{(} \frac{\sigma_{\pi N} }{2 m_\pi^2} 
+ \frac{ f_\pi^2 b^{+}}{2 m_\pi^2 } \Big{)} + \frac{g_A^2 k_F^4}{6 \pi^4 f_\pi^4} F(\frac{m_\pi}{2 k_F}) - \frac{k_F^4}{16 \pi^4 f_\pi^4}   \Big{]}.
\end{eqnarray}
The LECs are determined by the $\pi N$ sigma term $\sigma_{\pi N}$ and 
the isoscalar $\pi N$ scattering length $b^+$. 

  \begin{figure}[t]
    \centering
    \includegraphics[width=6cm,angle=-90]{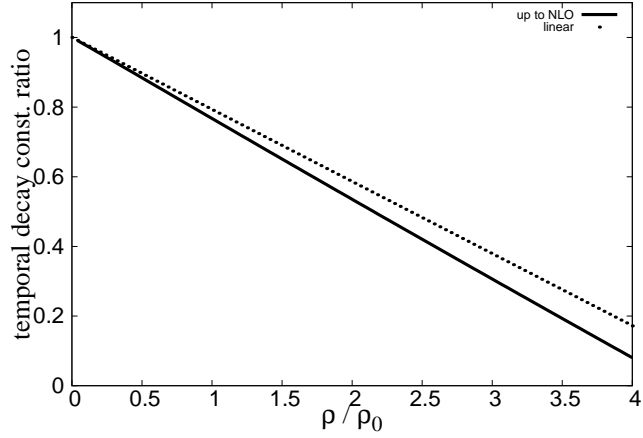}
    \caption{Density dependence of the temporal decay constant normalized by in-vacuum pion decay constant $f_\pi$ in symmetric nuclear matter.
             The dotted line is for the result in the leading linear density and 
             the solid line is for the result including the next-to-leading order.}
    \label{decay_dep}
  \end{figure}

We show in Fig.\ref{decay_dep} the density dependence of the temporal decay constant.
In this figure, the dotted line is for the result of the leading linear density 
and the solid line is for the result including the next-to-leading order $O(k_F^4)$.
We find that NLO correction gives a few percents change in low density region but in high density region about three times the normal nuclear density NLO correction contributes 10 percents and not small.

Let us make some comments on the perturbative expansion in terms of the Fermi momentum.
In the in-medium CHPT, the expansion parameter is to be $ k_F / (4 \pi f_\pi) $ and it amounts to around $ 0.3 $ to $ 0.4 $ even at higher density region than the saturation density.
With this small value, we could say that the perturbative expansion in terms of the Fermi momentum may have good convergence. Nevertheless, if one wants to judge whether the Fermi momentum expansion is good or not in nuclear matter, one should describe the nuclear matter and reproduce the saturation properties first, and see the higher density effects.
The nuclear matter could not be described essentially in perturbation theory of the in-vacuum fields.
One could need nonperturbative treatment to obtain the nuclear matter.
In such a case, naive perturbative expansion in terms of the Fermi momentum would break down before the nuclear density.

\subsection{In-medium pseudo-scalar coupling}

\begin{figure}[t]
  \begin{center}
  \includegraphics[ width=10cm ]{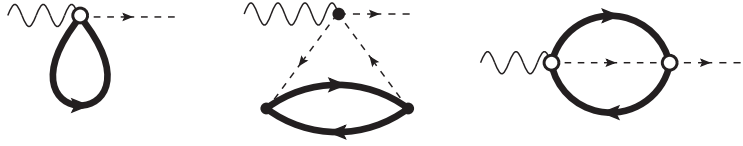}
  \caption{Feynman diagrams contributing to the one-particle irreducible
  vertex correction for the pseudo-scalar coupling $\hat{G}_\pi $. 
  The wavy line is the pseudo-scalar density.
  The left diagram is the leading order density contribution, while 
  the middle and right diagrams contribute to the next-to-leading order correction.} 
           \label{fig:gpi}
  \end{center}
\end{figure}

The pseudo-scalar coupling $G_\pi^{*}$ is also calculated as
\begin{equation}
G_\pi^* \equiv \langle \Omega | P^a | \pi^{*b} \rangle = \hat{G}_\pi \sqrt{Z} .
\end{equation}
In Fig.~\ref{fig:gpi}, we list up the Feynman graphs which contribute to the 
one-particle irreducible vertex correction $\hat G_{\pi}$.
Performing the same calculation as the pion decay constant $\hat f_{t}$,
we obtain the density dependence of the in-medium pseudo-scalar coupling:
\begin{eqnarray}
\frac{G_\pi^*}{G_\pi}
&=& 1 + \frac{\rho}{f^2} \big{(} 4c_1 - c_2 - c_3 + \frac{ g_A^2}{8 m_N} \big{)} +  \frac{g_A^2 k_F^4}{12 f^4 \pi^4} F(\frac{k_F}{2 m_\pi}) \nonumber \\
&&+ \frac{ k_F^4}{f^4 \pi^4} \Big{(} \frac{1}{16} + m_\pi^2 (2c_1 -c_2 -c_3)^2 \Big{)}  .
\end{eqnarray}

\subsection{In-medium low energy theorems}

It is known that the Gell-Mann--Oakes--Renner (GOR) relation is satisfied 
in medium up to the linear density (see, for example Ref.~\cite{Jid08}). 
Having obtained the density dependences of the pion decay constant and 
the pion mass in medium up to $O(k_{F}^{4})$, we are going to
examine the GOR relation satisfied at $O(k_{F}^{4})$. The density 
dependence of the quark condensate is obtained in Ref.~\cite{God13}
up to $O(k_{F}^{4})$. Using the expressions of the in-medium quantities,
we obtain
\begin{eqnarray}
\frac{f_t^2}{f_\pi^2} \frac{m_\pi^{*2}}{m_\pi^2}
-\frac{\langle \bar{u}u + \bar{d}d \rangle^*}{\langle \bar{u}u + \bar{d}d \rangle_0} 
&=& \frac{8 m_\pi^2 k_F^4}{3 \pi^4 f^4} \Big{[} (2c_1 - c_2 - c_3 )^2 G(0) - 4c_1^2 G_1 (\frac{m_\pi^2}{4k_F^2}) \Big{]} \\
G_1 (x^2) = G( -x^2 ) &=& \frac{3}{8} - \frac{x^2}{4} -x \arctan \frac{1}{x} + \frac{x^2}{4} (x^2 +3 ) \ln |\frac{1+x^2}{x^2}|. 
\label{GOR}
\end{eqnarray}
From this equation, we find that the GOR relation is not satisfied at this order
and the double scattering terms break the relation.  
The reason of the breaking is the following:
The GOR relation connects the chiral condensate and the pion quantities.
The chiral condensate is calculated in the soft limit by taking $q\to 0$
in the correlation function, while the pionic quantities  
are evaluated for pion on the mass shell. Off the chiral limit, where pion is massive,
the two energy points are different. This difference makes the GOR relation broken.

In the chiral limit in vacuum, one has the Glashow--Weinberg (GW) relation 
$f_{\pi}G_{\pi} = -\langle \bar qq\rangle$. This relation is known to be
satisfied in medium in the chiral limit and at the linear density~\cite{Jid08}.
Here we also directly check whether the relation is satisfied 
in the order $O(k_{F}^{4})$ off the chiral limit:
\begin{equation}
\frac{f_t^{*}}{f_\pi} \frac{G_\pi^*}{G_\pi} - \frac{\langle \bar{u}u + \bar{d}d \rangle^*}{\langle \bar{u}u + \bar{d}d \rangle_0}
= \frac{8 m_\pi^2 k_F^4}{3 \pi^4 f^4} \Big{[} (2c_1 - c_2 - c_3 )^2 G(0) - 4c_1^2 G_1 (\frac{m_\pi^2}{4k_F^2}) \Big{]}.
\label{GWrel}
\end{equation}
As seen in this equation, the GW relation is broken by the double scattering term and the breaking terms correspond to the right-hand side in Eq.(\ref{GOR}).

It is also interesting to discuss the in-medium sum rule 
for the in-medium quark condensate derived in Ref.~\cite{Jid08} model-independently
in the chiral limit:
\begin{equation}
- \langle \bar{q}q \rangle = {\rm Re} \sum_\alpha f_\alpha G_\alpha , 
\end{equation}
where $\alpha$ is the label of the zero-modes in nuclear matter, such as
the pions and particle-hole excitations, and summation is taken over all of the zero modes. 
It has been shown that, in the linear density, only the pion zero mode
can contribute to the sum rule and one obtains the in-medium Glashow--Weinberg
relation $f_t G_{\pi}^* = -\langle \bar qq\rangle^*$. The other zero modes,
such as particle-hole excitations, contributes only as higher density correction.
As discussed in Ref.~\cite{God13}, the $NN$ correlation effects come into the
chiral effective theory from the order of $O(\rho^{2})$. Thus, we expect 
that the pion mode dominates the sum rule beyond the linear density 
but below the square density order in the chiral limit. We can check it
by taking the chiral limit in Eq.~\eqref{GWrel}. This means that the GW relation 
is satisfied up to $O(k_{F}^{4})$ in the chiral limit:
\begin{equation}
   f_{t} G_{\pi}^* = -\langle \bar uu + \bar dd \rangle.
\end{equation}
With this relation,  realizing that both $f_{t}$ and $G_{\pi}^{*}$ decrease 
as the density increases, we find that the chiral condensate decreases
more rapidly than the pion decay constant. Although one might have
some corrections from $O(\rho^{2})$ contributions, 
this observation could be a hint that at a certain higher density
the chiral condensate would become zero before the pion decay 
constant would be zero. In this situation,  even though the quark condensate
is zero the chiral symmetry is spontaneously broken.

\subsection{In-medium $\pi^0 $ decay to $2\gamma$}
The neutral $\pi^{0}$ meson decays dominantly into $2 \gamma$ in vacuum through 
the quantum effect known as the axial anomaly. Here let us estimate the 
medium modification of the decay rate of $\pi^{0}$ to $2\gamma$ by following
the formulation presented in Sec.\ref{decayconst}. The decay amplitude may be
written, as requested by Lorentz invariance, in the form
\begin{equation}
  \langle \pi^{0*} | \gamma \gamma \rangle 
  = - i M^{*}_{\gamma\gamma} \epsilon^{\mu\nu\alpha\beta} 
  \epsilon^{*}_{\mu}k_{\nu} \epsilon_{\alpha}^{\prime *} k^{\prime}_{\beta}, 
\end{equation}
where $M^{*}_{\gamma\gamma}$ stands for the decay amplitude in medium,
$\epsilon^{\mu\nu\alpha\beta}$ is the totally antisymmetric tensor,
$\epsilon^{*}_{\mu}$ and $\epsilon^{*\prime}_{\alpha}$ are the polarization 
vectors of the emitted photons, and $k_{\nu}$ and $k_{\beta}^{\prime}$
are the momenta of the photon. 
According to the argument given in Sec.\ref{decayconst}, the in-medium 
amplitude $M_{\gamma\gamma}^{*}$ can be written as
\begin{equation}
  M^{*}_{\gamma\gamma} = \sqrt{Z} \hat{M}_{\gamma\gamma}
\end{equation}
with the wave function renormalization $Z$ and the medium correction 
of the one-particle irreducible $\pi^0 \gamma \gamma$ vertex $\hat {M}_{\gamma\gamma}$.

In the nuclear medium, the $\pi^{0}$ meson may decay into $2\gamma$ through
non-anomalous diagrams because of the presence of nucleons, which can bring 
out intrinsic parity violated in the $\pi^{0} \to 2\gamma$ process. Nevertheless,
in Ref.~\cite{Mei02}, it has been found that there are no medium corrections
to the $\pi^{0} \to 2\gamma$ amplitude up to ${\cal O}(p^{5})$, that is,
the linear density correction in the density expansion. The argument is based
on the spin and flavor structure of the vertices. The nucleon one loop with 
the $\pi^{0} \gamma \gamma NN$ contact vertex should vanish because
the contact vertex contains $\gamma_{5}$ and there are no other vertices 
can bring $\gamma_{5}$ in the nucleon loop, which makes  the trace of
the Dirac matrices vanish.  The diagrams in which the photons and the neutral 
pion emit sequentially from a nucleon also turn out to vanish according 
to careful calculation of the flavor matrix. Hence, the in-medium correction
up to the linear density vanishes and the in-medium vertex correction 
$\hat{M}_{\gamma\gamma}$ amplitude is equivalent to the in-vacuum
amplitude $M_{\gamma\gamma}$. This leads to the conclusion that 
the in-medium correction of the decay amplitude ${M}_{\gamma\gamma}^{*}$
can be calculated solely by the wave function renormalization as 
\begin{equation}
   \frac{M_{\gamma\gamma}^{*}}{M_{\gamma\gamma}} = \sqrt Z .
\end{equation}

Ignoring the small density correction of the pion mass in the phase space
of the decay process, we find that the modification of the 
$\pi^{0}$ decay rate to $2\gamma$ in nuclear matter is given by 
the wave function renormalization as
\begin{equation}
\frac{\Gamma^*}{\Gamma_0} = Z .
\end{equation}
This implies that the $\pi^{0}$ decay into $2\gamma$ in nuclear matter
measures directly the in-medium renormalization of the $\pi^{0}$ wave function
in lower density. Using the result discussed in the previous subsections, 
we obtain in the linear density approximation as
\begin{equation}
\frac{\Gamma^*}{\Gamma_0} = 1+ 0.4 \frac{\rho}{\rho_0}.
\end{equation}
At normal nuclear density the decay rate is 1.4 times 
enhanced  and the partial width becomes about 10 eV.

\section{Summary}
\label{summary}
We have discussed the in-medium pion properties, such as the pion decay
constant, the pion mass and the wave function renormalization based on the
in-medium chiral perturbation theory. First, we have provided a general formalism
of the in-medium chiral perturbation theory and have discussed an expansion in terms
of Fermi momentum. Assuming that the renormalization for the in-vacuum quantities
is performed in an appropriate way, we use the observed values to determine
the low energy constants (LECs) in the chiral Lagrangian. Since we have used the
physical values, the higher order corrections for the momentum expansion are
implicitly included in the calculation. Thus, we focus on the expansion of
the Fermi momentum of the physical quantities, which are calculated by
the QCD current Green functions.
To calculate the in-medium quantities, we carefully define the in-medium
pion state, and we have found that the in-medium wave function renormalization
plays an essential role to define the in-medium coupling constants, such as
the decay constant and pseudo-scalar coupling constant.

We have evaluated the density dependence of the decay constant, the pion mass,
the pion wave function renormalization and the pseudo-scalar coupling including
the next-to-leading order of the density expansion beyond the well-known linear density
approximation based on the in-medium chiral perturbation theory.
We have found that the $O(k_F^4)$ corrections give a few percents changes
in the low density region for the decay constant and the pion mass, while
in higher density such as three times saturation density, the corrections
can be as the order of 10 to 20 percents and can give significant contribution.
We have also found that the wave function renormalization is enhanced
as largely as 50 percents at the saturation density.
The main contribution among the corrections comes from the double scattering term.
In addition, we have checked whether the low energy theorems, the Gell-Mann--Oakes--Renner
relation and the Glashow--Weinberg relation are satisfied in medium beyond the linear density
approximation. We have found that these relations are not satisfied at $O(k_F^4)$
off the chiral limit. The origin of the breaking is that we use the different energy values
to evaluate the pion quantities and the chiral condensate; we take the soft limit
to obtain the chiral condensate, while we take the pion on shell point, that is $q_0=m_\pi^*$,
to evaluate the pion quantities. Finally, we have discussed the density dependence of
the $\pi^0 \to \gamma \gamma$ decay width. Considering the spinor and flavor vertex
structure of the chiral interactions, we have found that the density dependence of
the $\pi^0 $ width comes from the wave function renormalization alone at linear density
order. With this observation,  the wave function renormalization $Z$ would be measured
directly in the decay process.

\section*{Acknowledgement}
This work was partially supported by the Grants-in-Aid for Scientific Research (No.\ 25400254, No.\ 24105706 and No.\ 24540274).

\appendix

\section{Explicit expression of each interaction term}
\label{app:lagrangian}
In order to make the perturbative calculation simple, we use
the following parametrization of pion field $U$ given by 
Ref.~\cite{Charap:1971bn,Gerstein:1971fm}, 
as we have done in the previous paper~\cite{God13}:
\begin{equation}
   U = \exp\left[ i \pi^i \tau^i \frac{y(\pi^{2})}{2 \sqrt{\pi^{2}}} \right]
\end{equation}
where $y(\pi^{2})$ satisfies
\begin{equation}
   y - \sin y = \frac{4}{3} \left(\frac{\pi^{2}}{f^{2}} \right)^{\frac{3}{2}}.
\end{equation}

The expansion of the chiral field $U$ in terms of the pion field is
given in Ref.~\cite{Charap:1971bn} as
\begin{equation}
U= 1+ \frac{i \pi^i \tau^i}{f} - \frac{\pi^2}{2 f^2} - \frac{i \pi^i \tau^i \pi^2 }{10 f^3} - \frac{(\pi^2)^{2} }{40 f^4} + \cdots 
\end{equation}
and for the chiral field $u$
\begin{equation}
u= 1+ \frac{i \pi^i \tau^i}{2f} - \frac{\pi^2 }{8 f^2} + \frac{i \pi^i \tau^i \pi^2 }{80 f^3} - \frac{9 (\pi^2)^{2}  }{640 f^4} + \cdots .
\end{equation}
In this parametrization, the soft pion theorems, such as Adler's condition,
remain satisfied in simple perturbation theory.
In addition, we show chiral interactions which we use in the following calculation.

With this parametrization, one finds the explicit expression of each interaction 
term in the chiral Lagrangian. 
The chiral interactions for pions are the followings:
\begin{eqnarray}
\mathcal{L}^{(2)}_{\pi^4} &=& - \frac{1}{10 f^2} \partial_\mu \pi^i \partial^\mu \pi^j \pi^k \pi^l (\delta^{i j} \delta^{kl} - 3 \delta^{ik} \delta^{jl}  ) - \frac{m_q B_0}{20 f^2} \pi^i \pi^j \pi^k \pi^l \delta^{ij} \delta^{kl}, \\
\mathcal{L}^{(2)}_{\pi^3 a} &=& \frac{1}{5f} a_\mu^i \partial^\mu \pi^j \pi^k \pi^l (3\delta^{ij} \delta^{kl} - 4 \delta^{ik} \delta^{jl}) .
\end{eqnarray}
The interactions containing nucleon in the bilinear form read
\begin{eqnarray}
A_{\pi}^{(1)} &=& \frac{g_A}{2f} \gamma^{\mu} \gamma_5 \partial_{\mu} \phi^i \tau^i 
   \label{eq:piNN} \\
A_{a}^{(1)} &=& - g_A \gamma^\mu \gamma_5 a_\mu^i \frac{\tau^i}{2} \\
A_{\pi a}^{(1)} &=& \frac{i}{2f} \gamma^\mu [\phi, a_\mu ] = - \frac{1}{2f} \gamma^\mu \phi^i a_\mu^j \epsilon^{ijk} \tau^k \\
A_{\pi a}^{(2)} &=& -\frac{2c_2}{f m_N^2} \partial_{\mu} \phi^i a_{\nu}^i \partial^{\mu} \partial^{\nu} +\frac{2c_3}{f} \partial_{\mu} \phi^i a^{\mu i} .
\end{eqnarray}

\section{Details of self energy calculations}
\label{app:calc}
In this appendix, we show the details of the calculations of the self energy shown in Sec.\ref{results}.
We consider the isospin symmetric limit and the symmetric nuclear matter.
The external momentum is fixed as $q= (q_0, {\bf 0} )$ for simplicity.
We write the nucleon propagator in the Fermi sea as
\begin{equation}
i G_{m}(p) = - 2 \pi (\slash{p} + m_N) \delta (p^2 - m_N^2) \theta (p_0) \theta (k_F - |{\bf p}|).
\end{equation}
For the Pauli-blocked nucleon propagator in the 
symmetric nuclear matter $iG(p)$, we use the following expression, which
is equivalent to Eq.~\eqref{fermigas}:
\begin{eqnarray}
i G(p) &=& i \frac{ \slash{p}+m_N}{2 E({\bf p})} \left( \frac{1- \theta ( k_F- |{\bf p}| ) }{p_0 -E({\bf p}) +i \epsilon } 
 + \frac{\theta ( k_F - |{\bf p}| )}{p_0 -E({\bf p}) -i \epsilon } 
+ \frac{1}{p_0 +E({\bf p}) -i \epsilon }  \right) \\
&=& i \frac{ \slash{p}+m_N}{2 E({\bf p})} G_r(p) 
\end{eqnarray}
We also define the propagator $iG_r (p)$ in which the spinor structure and 
the nucleon energy are factored out from $iG(p)$. 
In the following calculation, the trace symbol ${\rm Tr}$
implies to take trace in both spinor and isospin spaces, while ${\rm Tr}_{s}$
means the trace for only the spinor space.

\subsection{1-loop integrals}
We first calculate the pion self energy given by the diagram in the left of Fig.\ref{self}(a):
\begin{eqnarray}
-i \Sigma_1 (q_0) \delta^{ij} &=& - \int \frac{d^4 p}{(2 \pi )^4} {\rm Tr} \Big{[} (-i A^{(1)}_\pi) i G(p+q) (-i A^{(1)}_\pi ) iG(p) \Big{]} \nonumber \\
&=& - \int \frac{d^4 p}{(2 \pi )^4} {\rm Tr} \Big{[} (-i) \frac{g_A}{2f} \gamma_5 i \slash{q} \tau^i i \frac{\slash{p}+\slash{q} +m_N }{2E({\bf p}+{\bf q})}  G_r (p+q) \nonumber \\
&& \phantom{- \int \frac{d^4 p}{(2 \pi )^4} {\rm Tr} \Big{[}} \times
(-i) \frac{g_A}{2f} (- \gamma_5 ) i \slash{q} \tau^j i \frac{\slash{p} +m_N }{2E({\bf p})}  G_r (p)  \Big{]} \nonumber \\
&=& - \delta^{ij} \frac{g_A^2}{2f^2 }  \int \frac{d^4 p}{(2 \pi )^4} {\rm Tr}_s \Big{[} \slash{q} (\slash{p}+\slash{q} -m_N) \slash{q} (\slash{p}+m_N) \Big{]} \frac{G_r (p+q) G_r (p)}{2E({\bf p}) 2E({\bf p+q})} .\ \ \ \ 
\end{eqnarray}
The spinor trace can be calculated as
\begin{eqnarray*}
f_{\rm Tr}(p_{0} ) = {\rm Tr}_s \Big{[} \slash{q} (\slash{p}+\slash{q} -m_N) \slash{q} (\slash{p}+m_N) \Big{]} 
= 4(p\cdot q)  \{ (p+q)^2 -p^2 \} -4q^2 (p^2 +m_N^2) .
\end{eqnarray*}
For the calculation of the integral, we perform the $p_0$ integral first using the Cauchy theorem along a contour of upper semicircle in 
the complex $p_0$-plane:
\begin{eqnarray}
&& \int \frac{d p^0}{2 \pi} f_{\rm Tr}(p_{0}) G_r(p+q) G_r(p) \nonumber \\
&=& 2i \Big{[}  f_{\rm Tr}(E({\bf p})-q_{0}) \frac{ (E({\bf p})-q_{0}) \theta (k_F - |{\bf p} | )  }{q_0(q_{0} -2E({\bf p} )) } + f_{\rm Tr}(E({\bf p})) \frac{ (E({\bf p}) + q_0 )\theta (k_F - |{\bf p}| ) }{q_{0}(q_0 + 2E({\bf p})) } \Big{]} \nonumber \\
&=& 8i q_{0}^{2} m_{N} \theta(k_{F}-|{\bf p}|) . \nonumber 
\end{eqnarray}
Here we have picked up only the terms including the step function, because the vacuum part should be subtracted for the calculation of the in-medium quantities, and in the final expression we have taken the leading term in the $1/m_N$ expansion.

Finally we obtain
\begin{eqnarray}
-i\Sigma_1 (q_0) \delta^{ij} 
= - \delta^{ij} \frac{i g_A^2 q_0^2 }{f^2 m_N}  \int \frac{d^3 p}{(2 \pi )^3} \theta (k_F - |{\bf p}| ) 
= - \frac{i g_A^2 q_0^2}{4 f^2 m_N} \delta^{ij} \rho ,
\end{eqnarray}
where we have used $ \rho = 2k_F^3 / (3 \pi^2)$.

Next we consider the self energy given by the right diagram in Fig.~\ref{self}(a):
\begin{eqnarray}
-i \Sigma_2 (q_0) &=& (-1) \int \frac{d^4 p}{(2\pi)^4} {\rm Tr} \Big{[} (-iA_{\pi \pi}^{(2)}) iG_m (p) \Big{]} \nonumber \\
&=& \int \frac{d^3 p}{(2 \pi )^3}  (-4i) \Big{(} \frac{8B_0 c_1 m_q}{f^2} -\frac{2c_2}{f^2 m_N^2} (q\cdot p)^2  - \frac{2c_3}{f^2} q^2 \Big{)} \theta (k_F - |{\bf p}| ) \nonumber \\
&=& - \frac{2 i \rho}{f^2} ( 4 c_1B_0 m_q  - (c_2 + c_3 ) q_0^2 )  .
\end{eqnarray}
Here we have taken the leading term of the $1/m_{N}$ expansion.

\subsection{2-loop integrals}
Next we calculate the next-to-leading contributions in the density expansion. 
Here we denote the pion propagator as $i D_\pi (p)$.

First we consider the left diagram in Fig.~\ref{self}(b) $\Sigma_3 (q_0) $:
With symmetric factor $1/2$, the contribution is written as
\begin{equation}
 -i \Sigma_3 (q_0) = \frac{1}{2} \int \frac{d^4 k}{(2\pi)^4} i {\mathcal L}_{\pi^4}^{(2)}  (i D_\pi (k) )^2 (-i\Sigma_m (k)) ,
\end{equation}
where, $\Sigma_m (k)$ is defined as
\begin{displaymath}
-i \Sigma_m (k) = - \int \frac{d^4 p}{(2 \pi )^4} {\rm Tr} \Big{[} (-i A^{(1)}_\pi) i G_m(p+k) (-i A^{(1)}_\pi ) iG_m(p) \Big{]} .
\end{displaymath}
Using the on-shell conditions $(p+k)^2 = m_N^2$, $p^2 = m_N^2 $ for the 
nucleon propagators in the Fermi sea, the spinor trace is calculated as follows 
\begin{displaymath}
{\rm Tr_s} \Big{[} \slash{k} (\slash{p} + \slash{k} - m_N ) \slash{k}  (\slash{p} + m_N )  \Big{]} = -8 m_N^2 k^2 .
\end{displaymath}
Hence, we obtain $\Sigma (q_0)$:
\begin{eqnarray}
-i \Sigma_3 (q_0) 
&=& \frac{i g_A^2 m_\pi^2 }{2 f^4} \int \frac{d^3 k}{(2\pi)^3} \frac{d^3 p}{(2 \pi )^3} (i D_\pi (k) )^2  k^2 \theta (k_f - |{\bf p}+ {\bf k}|) \theta (k_f - {\bf p}) \nonumber \\
&=& \frac{i g_A^2 m_\pi^2}{2 f^4} \int \frac{d^3 k}{(2\pi)^3} \frac{d^3 p}{(2 \pi )^3} \frac{ {\bf k}^2 }{({\bf k}^2 + m_\pi^2 )^2} \theta (k_f - |{\bf p}+ {\bf k}|) \theta (k_f - {\bf p}) \nonumber \\
&=& \frac{i g_A^2 m_\pi^2 k_F^4}{12 \pi^4 f^4} F( \frac{m_\pi^2}{4 k_F^2} ) ,
\end{eqnarray}
where we have used the following integral formula:
\begin{eqnarray*}
&& \int \frac{d^3 k}{(2 \pi )^3} \frac{d^3 p}{(2 \pi )^3} \frac{{\bf k}^2}{ ({\bf k}^2 + m_{\pi}^2)^2} \theta (k_F -|{\bf p}+\frac{{\bf k}}{2}|) \theta (k_F - |{\bf p}-\frac{{\bf k}}{2}|) = \frac{k_F^4}{6 \pi^4} F( \frac{m_\pi^2}{4 k_F^2} ),
\end{eqnarray*}
with function $F(a^2)$:
\begin{eqnarray*}
F(a^2)&=&\int_{0}^{1}dx \left(\frac{x^{2}}{x^{2}+a^{2}}\right)^{2} \frac{1}{2} (1-x)^{2} (x+2)\\
&=& \frac{3}{8} -\frac{3a^2}{4} -\frac{3a}{2} {\rm arctan} \frac{1}{a}
 + \frac{3a^2}{4} (a^2 +2) \ln \frac{a^2 +1}{a^2} .
\end{eqnarray*}

We consider the contribution coming from the middle diagram in Fig.~\ref{self}(b), $\Sigma_4 (q_0)$:
\begin{eqnarray*}
-i\Sigma_4 (q_0) &=& - \int \frac{d^4 p}{(2 \pi )^4} \frac{d^4 k}{(2 \pi )^4} {\rm Tr} \Big{[} (-i A^{(1)}_{\pi \pi}) i G_m (k- \frac{p}{2} ) (-i A^{(1)}_{\pi \pi} ) iG_m (k+ \frac{p}{2}) iD_\pi (q-p) \Big{]} \\
&=& - \int \frac{d^4 p}{(2 \pi )^4} \frac{d^4 k}{(2 \pi )^4} {\rm Tr} \Big{[} \frac{\epsilon^{ikm} \tau^m}{4f^2} ( 2\slash{q} - \slash{p} )
i G_m (k- \frac{p}{2}) \frac{\epsilon^{kjm} \tau^m}{4f^2} ( 2\slash{q} - \slash{p} )  iG_m (k+ \frac{p}{2}) \Big{]} \\
\end{eqnarray*}
Performing integration in terms of $p_{0}$ and $k_{0}$,
we obtain 
\begin{eqnarray}
k_0 &=& \frac{1}{2} (E( {\bf k} +{\bf p}/2) + E({\bf k}-{\bf p}/2) ) \\
p_0 &=& E({\bf k}+{\bf p}/2) - E({\bf k}-{\bf p}/2)
\end{eqnarray}
which provide also $k\cdot p = 0$, $k^2 + p^2 /4 = m_N^2$.
With these expression, the spinor trace is calculated as
\begin{eqnarray*}
 {\rm Tr}_s \Big{[} ( 2\slash{q} - \slash{p} ) ( \slash{k} - \frac{ \slash{p}}{2} + m_N )( 2\slash{q} - \slash{p} ) ( \slash{k} + \frac{ \slash{p}}{2} + m_N ) \Big{]} 
= 32 (q\cdot k)^2 - 8 (q \cdot p)^2 + 8 q^2 p^2 
\end{eqnarray*}
In the leading term of the $1/m_N$ expansion, we have the relation $k_0 = m_N$, $p_0 = 0$ and we obtain
\begin{eqnarray*}
-i\Sigma_4 (q_0) 
& = & -\int \frac{d^3 p}{(2 \pi )^3} \frac{d^3 k}{(2 \pi )^3} \frac{2 q_0^2 }{f^4} \theta (k_f - |{\bf k} - \frac{{\bf p}}{2}|) \theta (k_f - |{\bf k} +  \frac{{\bf p}}{2} |) \frac{i}{{\bf p}^2 -( q_0^2 - m_\pi^2)} .
\end{eqnarray*}
Using the following integral formula~\cite{Fet03}
\begin{eqnarray*}
&& \int \frac{d^3 k}{(2 \pi )^3} \theta (k_f - |{\bf k} - \frac{{\bf p}}{2}|) \theta (k_f - |{\bf k} +  \frac{{\bf p}}{2} |) = \frac{k_F^3}{6 \pi^2} (1-\frac{3}{2} x + \frac{1}{2} x^3) \theta (1-x) ,
\end{eqnarray*}
with $x= |{\bf p}|/(2k_F)$,
we get 
\begin{eqnarray}
-i \Sigma_4 (q_0) 
&=& - \frac{i q_0^2 k_F^4}{3 \pi^4 f^4} G( \frac{q_0^2 - m_\pi^2}{4 k_F^2}    )  ,
\end{eqnarray}
where we have used
\begin{displaymath}
G( a^2 ) = \int_{0}^{1} \frac{x^{2}}{x^{2}-a^{2}} (1-x)^{2}(x+2) dx =
\frac{3}{8} + \frac{a^2}{4}  + \frac{a}{2} \ln | \frac{1- a}{1+ a} | + \frac{a^2}{4} (a^2 -3) \ln | \frac{1-a^2}{a^2} | .
\end{displaymath}

Finally we consider $\Sigma_5 (q_0)$ given by the right diagram in Fig.~\ref{self}(b).
This graph corresponds to the double scattering correction called as Ericson-Ericson term \cite{Eri66} and is calculated as:
\begin{eqnarray*}
-i \Sigma_5 (q_0) &=& - \int \frac{d^4 p}{(2 \pi )^4} \frac{d^4 k}{(2 \pi )^4} {\rm Tr} \Big{[} (-i A^{(2)}_{\pi \pi}) i G_m (k- \frac{p}{2} ) (-i A^{(2)}_{\pi \pi} ) iG_m (k+ \frac{p}{2}) iD_\pi (p+q) \Big{]} \\
&=& \int \frac{d^3 p}{(2 \pi )^3} \frac{d^3 k}{(2 \pi )^3}
{\rm Tr}_s \Big{[} (\slash{k} - \frac{\slash{p}}{2} + m_N ) (\slash{k} + \frac{\slash{p}}{2} + m_N ) \Big{]} \frac{ \delta^{ij}}{2m_N^2} iD_\pi (p+q) \\
&&\times \Big{(} \frac{8c_1 B_0 m_q }{f^2} - \frac{2c_2}{f^2 m_N^2} \{ q\cdot (k- \frac{p}{2}) \} \{ (p+q)\cdot (k-\frac{p}{2} ) - \frac{2c_3}{f^2} q \cdot (p+q) \} \Big{)} \\
&& \times  \Big{(} \frac{8c_1 B_0 m_q }{f^2} - \frac{2c_2}{f^2 m_N^2} \{ q\cdot (k+ \frac{p}{2}) \} \{ (p+q)\cdot (k+\frac{p}{2} ) - \frac{2c_3}{f^2} q \cdot (p+q) \} \Big{)} \\
&& \times \theta (k_f - |{\bf k} - \frac{{\bf p}}{2}|) \theta (k_f - |{\bf k} +  \frac{{\bf p}}{2} |).
\end{eqnarray*}
The spinor trace is reduced to
\begin{eqnarray*}
{\rm Tr}_s \Big{[} (\slash{k} - \frac{\slash{p}}{2} + m_N ) (\slash{k} + \frac{\slash{p}}{2} + m_N ) \Big{]}
&=& 4 k^2 - p^2 +m_N^2 = 8m_N^2 - 2 p^2
\end{eqnarray*}
Finally $ \Sigma_5 (q_0) $ is obtained as
\begin{eqnarray}
-i\Sigma_5 (q_0) 
&= & - \frac{ 4i}{f^4} \Big{(} 8c_1 B_0 m_q  - 2c_2 q_0^2 - 2c_3 q_0^2 \Big{)}^2 \int \frac{d^3 p}{(2 \pi )^3} \frac{d^3 k}{(2 \pi )^3} \nonumber \\
&& \times
\theta (k_f - |{\bf k} - \frac{{\bf p}}{2}|) \theta (k_f - |{\bf k} +  \frac{{\bf p}}{2} |) \frac{1}{ {\bf p}^2 -(  q^2 - m_\pi^2 ) -i \epsilon } \nonumber \\
& = & - \frac{ 4i}{f^4} \Big{(} 8c_1 B_0 m_q  - 2c_2 q_0^2 - 2c_3 q_0^2 \Big{)}^2 \frac{k_F^4}{6 \pi^4} G(\frac{q^2 - m_\pi^2}{4 k_F^2}).
\end{eqnarray}
In the last equality, we have performed the integral in the same way as $\Sigma_{4}$.

\section{Singularity in derivative of double scattering term}
\label{infra}
The loop integral of the double scattering term reads
\begin{eqnarray}
   I_{\rm ds}(q_0) &=& \int \frac{d^{3}p}{(2\pi)^{3}} \frac{1}{q_0^{2} - m_{\pi}^{2} - {\bf p}^{2} + i\epsilon} 
   \int \frac{d^{3}k}{(2\pi)^{3}} \theta(k_{f} - | {\bf k} - \frac{{\bf p}}{2}|)
   \theta(k_{f} - | {\bf k} + \frac{{\bf p}}{2}|) \nonumber
\end{eqnarray}
where $q_0$ is the energy of the external line and we have taken ${\bf q} = 0$ for the external momentum.
Performing the following integral 
\begin{eqnarray}
\int \frac{d^{3}k}{(2\pi)^{3}} \theta(k_{f} - | {\bf k} - \frac{{\bf p}}{2}|)
    \theta(k_{f} - | {\bf k} + \frac{{\bf p}}{2}|)  = \frac{k_{F}^{4}}{6\pi^{2}} \left(1- \frac{3}{2} x + \frac{1}{2} x^{3}\right) \theta(1-x) ,
\end{eqnarray}
where $x = |{\bf p}|/(2k_{F})$, we write the loop integral $I_{\rm ds}(k_0)$ as
\begin{eqnarray}
   I_{\rm ds}(q_0) =
   -\frac{k_{F}^{4}}{6\pi^{4}} \int_{0}^{1} dx 
   \left(1- \frac{3}{2} x + \frac{1}{2} x^{3}\right)
   \frac{x^{2}}{x^{2} - a - i\epsilon} \label{eq:intds}
\end{eqnarray}
where we have define $a \equiv (q_0^{2} - m_{\pi}^{2})/(4k_{F}^{2})$. 
The integral with respect to $x$ can be done straightforwardly and we obtain
\begin{eqnarray}
   I_{\rm ds}(q_0) &=&
   -\frac{k_{F}^{4}}{6\pi^{4}}
   \left[ \frac{3}{8} + \frac{a}{4} 
   + \frac{\sqrt a}{2} \ln \left|\frac{1-\sqrt a}{1+\sqrt a}\right| 
   + \frac{a}{4}(a-3) \ln \left|\frac{1-a}{a} \right| \right]
\end{eqnarray}
This function is finite in the limit of $k_0 \to m_{\pi}$, {\it i.e.} $a \to 0$.

But the derivative of $I_{\rm ds}(q_0)$ with respect to $q_0^2$ has a singularity at $a \to 0$. The derivative is obtained as
\begin{eqnarray}
  \frac{\partial I_{\rm ds}(q_0)}{\partial q_0^2}
  &=& -\frac{k_{F}^{2}}{24\pi^{4}} \frac{1}{4} \left[ - \frac{2}{1+\sqrt a} 
  + \frac{1}{2\sqrt a} \ln \left| \frac{1-\sqrt a}{1+\sqrt a} \right|  + (2a - 3) \ln \left| \frac{1-a}{a} \right| \right]
\end{eqnarray}
This function is logarithmically divergent at $a \to 0$. In addition this loop integral starts from $k_{F}^{2} \ln k_{F}$ in the expansion of small $k_{F}$.
This behavior is contradict with the low density expansion where the leading order should be $\rho \sim k_{F}^{3}$.
Consequently, the wave function renormalization could have such a strange density dependence.
Thus this contribution is pathologic.
We do not take the exact limit of $q_0 \to m_{\pi}$ in the evaluation.
Nevertheless, we have to deal with the singularity, because the wave function renormalization would have inconsistent density dependence with the low density expansion and we may make evaluation of the physical quantities very close to the singular point, where the results are numerically unreliable.

The origin of the singularity seen in the derivative of the loop function at $q_0 = m_{\pi}$ can be identified when one performs derivative of Eq.~\eqref{eq:intds} in terms of $q_0$ and takes $a=0$:
\begin{eqnarray}
   I_{\rm ds}(m_\pi) &=&
   -\frac{k_{F}^{2}}{14\pi^{4}} \int_{0}^{1} dx 
   \left(1- \frac{3}{2} x + \frac{1}{2} x^{3}\right)
   \frac{x^{2}}{(x^{2} - i\epsilon)^{2}} \nonumber
\end{eqnarray}
The integral gets divergent when the integrand takes $x=0$ at the end point of the integral.
This singularity is very similar with the infrared divergence in quantum field theory.
Such infrared divergence should be cancelled with other diagrams emitting soft pions, when one calculates scattering rates, not in the amplitude itself.
Thus, the singularity found in the derivative of the loop function should be cancelled with other terms, when one calculates cross sections.
Relying on the above argument, we simply drop the term which includes the infrared singularity in this work.

\end{document}